\documentclass[graybox]{svmult}
\usepackage{makeidx}         
\usepackage{graphicx}        
\usepackage{multicol}        
\usepackage[bottom]{footmisc}
\usepackage{newtxtext}       %
\usepackage[varvw]{newtxmath}       

\usepackage{slashed}
\usepackage{nicefrac}
\usepackage{amsmath}
\usepackage{bbm}
\usepackage{bbold}
\makeindex             

\begin{document}

\title*{Simple Supergravity}
\author{Gianguido Dall'Agata and Marco Zagermann}
\institute{Gianguido Dall'Agata \at Dip. di Fisica e Astronomia ``Galileo Galilei'', Universit\`a di Padova and INFN, Sezione di Padova, Via Marzolo 8, 35131 Padova, Italy, 
\and Marco Zagermann}

\maketitle

\abstract{We present a short overview of the structure and couplings of supergravity theories at the component level. We do so with as little technical machinery as possible, working directly with the physical on-shell fields and using explicit computations and geometrical reasoning to arrive at the result, highlighting the new properties of supersymmetry in the context of a gravitational theory.}

\section{Introduction}

Supergravity is soon going to turn 50 \cite{Freedman:1976xh,Deser:1976eh}.
During this half of a century it lived several lives and it has been used and studied from various different vantage points, including the analysis of quantum gravity and black hole physics, string theory, particle physics phenomenology, cosmology and mathematics.
Each of these different approaches advanced our understanding of the features and structure of supergravity theories and, in turn, supergravity brought new ideas and fertilized each of these fields of study.

One of the main problems of the uninitiated who is interested in supergravity is that it is a rather technical subject and most of the introductory books and reviews deal with it by first emphasizing some specific mathematical formalism (like superspace and superfields or the group manifold approach, or the superconformal approach) and only after some significant effort by the reader they enter into the discussion of the physical properties of the theory.
While each of these approaches has its advantages and can be at some point necessary to obtain significant progress, we felt that a more simple hands on introduction, where every aspect is dealt with directly at the component level, emphasizing the physical features and their mathematical origin, was missing.
For this reason, while there are already several great reviews that use one of the aforementioned techniques (see for instance \cite{VanNieuwenhuizen:1981ae,Nilles:1983ge,deWit:1985aq,Castellani:1991et,Castellani:1991eu,Wess:1992cp,VanProeyen:1999ni,Derendinger:2000xx,deWit:2002vz,VanProeyen:2003zj}), we worked on a new physics first introduction to the subject, which took shape in the lecture notes \cite{DallAgata:2021uvl}.

The current chapter is a short redacted excerpt of the more detailed and complete presentation given in \cite{DallAgata:2021uvl}.
Here we mainly focus on the very basic ingredients that are needed for a first introduction to the subject, which we hope will work as an invitation for the reader to deepen their knowledge of the subject.

We also stress that in this collection the reader is also going to find short introductions to some of the alternative approaches mentioned above, as well as various applications.

\section{What is supergravity?}
Depending on which aspect one wants to emphasize, one could define supergravity theories in three different ways:
\begin{enumerate}
\item Supergravity theories are supersymmetric field theories with gravity where the supersymmetry transformations act nontrivially also on the gravitational field. \item Supergravity theories are supersymmetric field theories in which supersymmetry is realized not only as a \emph{global} (\emph{rigid}) symmetry but as a \emph{local} (\emph{gauge}) symmetry.
\item Supergravity theories are field theories \footnote{We assume finitely many fields and couplings as well as consistency with unitarity and diverse spacetime backgrounds.} with consistently interacting spin-3/2 fields \footnote{For the sake of readability, we do not distinguish carefully here between spin and helicity, i.e.~``spin $s$'' should be understood as ``helicity $\pm s$'' in the massless case.}. In supergravity, these spin-3/2 fields are called gravitino fields (or gravitini).
\end{enumerate}   	
In this section, we illustrate why these three apparently different characterizations describe essentially the same class of theories.  
We do so in the assumption that gravity is described by Einstein's general theory of relativity, so that supergravity actions consist of the Einstein--Hilbert term plus a restricted class of matter actions coupled to gravity.  
To this end, we revisit the simplest \emph{globally} supersymmetric field theory in four dimensions, the free massless Wess--Zumino model for one chiral multiplet, and discuss how this theory has to be changed when supersymmetry is turned into a \emph{local} symmetry, following Definition 2. As we will see,
making supersymmetry local by a simple iterative procedure (the ``Noether method'') directly exhibits the need for the gravitino field (cf. Definition 3) and its superpartner, the graviton (cf. Definition 1), and suggests the supersymmetry transformation laws of these fields. 
We end this section with a discussion of some basic properties of the gravitino field.

\subsection{Promoting supersymmetry to a local symmetry\protect\footnote{This section is reprinted from  \cite{DallAgata:2021uvl} © 2021 Springer-Verlag GmbH Germany, part of Springer Nature. Reproduced with permissions. All rights reserved.}} 
\label{sec:promoting_supersymmetry_to_a_local_symmetry}

Consider the free massless  \index{Wess--Zumino model}Wess--Zumino model for one chiral multiplet $(\phi,\chi)$\index{chiral multiplet}, where $\phi(x)$ is a complex scalar and $\chi(x)$ a Majorana spinor field\footnote{Throughout this article we use anti-commuting four-component Majorana spinors to describe fermionic degrees of freedom. Our conventions are summarized in Appendix \ref{sec:spinors} and follow the textbook \cite{DallAgata:2021uvl}, where many further details can be found.
} with Lagrangian 
\begin{equation}
	{\cal L} = - \partial_\mu \phi \partial^\mu \phi^* - \left(\overline \chi_R \slashed{\partial} \chi_L + \overline \chi_L \slashed{\partial} \chi_R \right) . 
	\label{WZLagrangian} 
\end{equation}
We recall that the mass dimensions of these fields are $D[\phi]=1$, and $D[\chi]= \nicefrac32$. 

The Lagrangian (\ref{WZLagrangian}) is invariant up to a total derivative under supersymmetric under the variations\footnote{Note that, since we are not  using auxiliary fields, the supersymmetry
algebra closes only on-shell:
$$\begin{array}{rcl}
[ \delta_{\epsilon_{2}} , \delta_{\epsilon_{1}} ]\phi &=&
\frac12(\overline{\epsilon}_{1}\gamma^{\mu}\epsilon_{2})\partial_{\mu}\phi   \\[2mm]
\left[ \delta_{\epsilon_{2}} , \delta_{\epsilon_{1}} \right]
 \chi_{L}  
&=&
  \frac12(\overline{\epsilon}_{1}\gamma^{\mu}\epsilon_{2})\partial_{\mu}\chi_{L}
  + [ \ldots ]\slashed{\partial}\chi_{L},
\end{array}
$$
where $[\ldots]$ denotes a non-vanishing expression of the fields and supersymmetry parameters. The last term then vanishes due to the field equation $\slashed{\partial}\chi_{L}=0$, and one obtains the usual susy algebra $\left[ \delta_{\epsilon_{2}} , \delta_{\epsilon_{1}} \right]
 =
  \frac12(\overline{\epsilon}_{1}\gamma^{\mu}\epsilon_{2})\partial_{\mu}$ on all fields.}
\begin{eqnarray}
	\delta_\epsilon \phi & = & \overline \epsilon_L \chi_L \quad\quad \Longleftrightarrow \quad \delta_\epsilon \phi^\ast \, = \, \overline \epsilon_R \chi_R \label{delphi} \\
	\delta_\epsilon \chi_L &=& \frac12 \slashed{\partial} \phi \epsilon_R \quad\,\Longleftrightarrow \quad\delta_\epsilon \chi_R = \,\frac12 \slashed{\partial} \phi^\ast \epsilon_L. \label{delchi} 
\end{eqnarray}
Note that $D[\epsilon] = -1/2$. 
In our conventions, (\ref{delchi}) is equivalent to 
\begin{equation}
	\delta_\epsilon \overline \chi_L = -\frac12 \overline \epsilon_R \slashed{\partial} \phi \quad\Longleftrightarrow\quad \delta_\epsilon \overline \chi_R= -\frac12 \overline \epsilon_L \slashed{\partial} \phi^\ast.\label{delchibar}
\end{equation}

To check this explicitly, we write the fermionic term of the Lagrangian (\ref{WZLagrangian}) as 
${\cal L}_{\textrm{fer}}= - \overline{\chi}_R\slashed{\partial}\chi_L + \partial_{\mu}(\overline{\chi}_{R})\gamma^\mu \chi_L$ and trace the terms involving $\epsilon_L$, which come only from the variation of $\phi$ and $\overline{\chi}_R$, because those proportional to $\epsilon_R$ follow by hermitean conjugation:
\begin{equation}
\delta {\cal L} =-\partial_{\mu}(\delta \phi)\partial^{\mu}\phi^\ast  -\delta \overline{\chi}_R\slashed{\partial}\chi_L + \partial_{\mu}(\delta \overline{\chi}_{R})\gamma^\mu \chi_L + \textrm{h.c.}
\end{equation}
Integrating by parts the first and the second term gives, using (\ref{delphi}), (\ref{delchi}) and (\ref{Clifford}),
\begin{eqnarray}
\delta {\cal L} &=&\delta \phi \Box \phi^\ast + 2 \partial_{\mu} (\delta \overline{\chi}_R)\gamma^{\mu}\chi_L+ 
\partial_{\mu}\underbrace{\Big( -\delta \phi\partial^\mu\phi^\ast - \delta \overline{\chi}_{R}\gamma^\mu\chi_L\Big) }_{\equiv\mathcal{K}^\mu}+\textrm{h.c.} \nonumber\\
&=& -\partial_\mu(\overline{\epsilon}_L)\slashed{\partial}\phi^\ast\gamma^{\mu}\chi_L
+ \partial_{\mu}\mathcal{K}^\mu + \textrm{h.c.}
\end{eqnarray}
As promised, the result is that under global supersymmetry, where the supersymmetry parameter is constant, $\partial_\mu \epsilon = 0$, the Lagrangian transforms into a total derivative:
\begin{equation}
	\delta_\epsilon {\cal L} = \partial_\mu ({\cal K}^\mu + {\cal K}^\mu{}^*) \equiv \partial_\mu K^\mu\,.
\end{equation}
 
When dealing with local supersymmetry, however, the parameter $\epsilon$  becomes a local function of the coordinates, $\epsilon = \epsilon(x)$, and the Lagrangian is no longer invariant up to a total derivative.
The new non-invariant part of the Lagrangian is
\begin{equation}
	\delta_\epsilon {\cal L}_{\textrm{new}} = (\partial_\mu \overline \epsilon) j^\mu = (\partial_\mu\overline \epsilon_L) j_L^\mu+(\partial_\mu \overline \epsilon_R) j_R^\mu, 
	\label{noninvar} 
\end{equation}
where
\begin{equation}
	j^\mu_{L}\equiv -\slashed{\partial} \phi^*  \gamma^\mu \chi_L,\qquad
	j^\mu_{R}\equiv -\slashed{\partial} \phi  \gamma^\mu \chi_R \label{curr} 
\end{equation}
give the super-Noether current $j^\mu = j^\mu_L + j^\mu_R$.
In fact, it can be easily checked that this supercurrent is a conserved current, namely that  $\partial_\mu j^\mu = 0$, upon using the equations of motion for the fields $\phi$ and $\chi$.
It should also be noted that the dimension of these currents is $D[j^\mu_{L,R}] = \nicefrac72$. 

We can now apply Noether's method\index{Noether's method} and associate to the supercurrent (\ref{curr}) a gauge field that compensates the non-invariance of the Lagrangian (\ref{noninvar}). 
This gauge field, $\psi_{\mu\alpha}$, has to have a spinorial index (i.e.~the index $\alpha=1,2,3,4$, which we will suppress again in the following) and a spacetime index ($\mu=0,1,2,3$), such that 
\begin{equation}
	\delta_\epsilon \psi_{\mu\, L,R} = M_P \partial_\mu \epsilon_{L,R}, \qquad	\delta_\epsilon \overline{\psi}_{\mu\, L,R} = M_P \partial_\mu \overline \epsilon_{L,R}, \label{deltapsi00}
\end{equation}
where $M_P$ is a mass parameter that is needed to relate the mass dimension $\nicefrac32$ of the fermionic field  $\psi_\mu$ and the dimension of the supersymmetry parameter $D[\epsilon] = - \nicefrac12$. As suggested by the notation, $M_P$ will later be identified with the (reduced) Planck mass.

The Noether procedure tells us that we need to add a new piece to the Lagrangian: 
\begin{equation}
	{\cal L}^\prime_{\textrm{WZ}} = - \frac{1}{M_P} \left(\overline \psi_{\mu L} j^\mu_L +\overline\psi_{\mu R} j^\mu_R\right) .\label{L'WZ} 
\end{equation}
Again $M_P$ is needed to get a Lagrangian density whose total mass dimension is 4, and this dimensionful coupling in the action
can be viewed as a first sign that we eventually need gravity in local supersymmetry.

Using (\ref{deltapsi00}) in the variation of (\ref{L'WZ}), we now precisely compensate the variation of the original Wess--Zumino multiplet, but now there is a new piece to compensate in the variation of (\ref{L'WZ}) from $\delta_\epsilon j^\mu_{R,L}$, which is in general non-vanishing. To see this, it suffices to consider the variation of the term
$\overline{\psi}_{\mu L} j^\mu_{L}$ that is quadratic in the scalar fields. This term comes from the variation of $\chi_L$ inside $j^\mu_L$:
\begin{equation}
	 \overline \psi_{\mu L} \slashed{\partial} \phi^* \gamma^\mu \delta_\epsilon\chi_L = \frac12 \overline \psi_{\mu L} \gamma^\nu \gamma^\mu \gamma^\rho\epsilon_R \partial_\nu \phi^* \partial_\rho \phi=  \overline \psi_{\mu L} \gamma_\nu \epsilon_R T^{\mu\nu} + \ldots 
\end{equation}
where, using some gamma matrix algebra, 
\begin{equation}
	T^{\mu\nu} = \partial^{(\mu} \phi \partial^{\nu)} \phi^* - \frac12 \eta^{\mu\nu} (\partial_\sigma \phi \partial^\sigma \phi^*), 
\end{equation}
and the dots stand for terms involving $\gamma^{\nu\mu\rho}$. One can show that variations bilinear in $\chi$ likewise give the energy momentum tensor for the field $\chi$. 
So, 
\begin{equation}
	\delta{\cal L}^\prime_{\textrm{WZ}} \sim \frac{1}{M_P} \overline \epsilon \gamma_\mu \psi_\nu T^{\mu\nu} + \ldots \label{deltaL'WZ} 
\end{equation}
In order to cancel this term, we now introduce a new current which is a symmetric tensor $g_{\mu\nu}$ with transformation rule 
\begin{equation}
	\delta g_{\mu\nu} \sim \frac{1}{M_P} \overline \epsilon \gamma_{(\mu}\psi_{\nu)},
	\label{deltag} 
\end{equation}
and add a new piece to the Lagrangian with a coupling between the tensor field $g_{\mu\nu}$ and  the energy momentum tensor: 
\begin{equation}
	{\cal L}''_{\textrm{WZ}} \sim -  g_{\mu\nu} T^{\mu\nu} \label{L"} . 
\end{equation}

As only the spacetime metric can couple to the energy momentum tensor, local supersymmetry requires the coupling of the Wess--Zumino multiplet to gravity described by a dynamical spacetime metric, $g_{\mu\nu}$, and $\psi_\mu$ must be its superpartner, the gravitino, as follows from the transformation law (\ref{deltag}). 
As in ordinary gauge theories, one also adds kinetic terms for these new ``gauge'' fields, and we thus expect a final result of the form 
\begin{equation}
	\begin{array}{rcccc}
		{\cal L} &=&\underbrace{ {\cal L}_{\textrm{kin}}(\phi) + {\cal L}_{\textrm{kin}}(\chi)} &+& \underbrace{{\cal L}_{\textrm{int}}(\phi,\chi, g_{\mu\nu},\psi_\mu)} \\
		&& \mbox{\footnotesize ${\cal L}_{\textrm{WZ}}$}&& \mbox{\footnotesize ${\cal L}^\prime_{\textrm{WZ}}+{{\cal L}''_{\textrm{WZ}}}+\ldots$}\\
		&+& {\cal L}_{\textrm{kin}}(g_{\mu\nu}) + {\cal L}_{\textrm{kin}}(\psi_\mu) 
	\end{array}
	\label{Ltot} 
\end{equation}
where the dots indicate possible further interaction terms. 

We used the chiral multiplet to guess the supersymmetry transformation rules of the supergravity multiplet. 
These rules, however, should hold also in the absence of the chiral multiplet, and we thus arrive at a motivated guess for the Lagrangian and transformation laws of pure ${\cal N} = 1$ supergravity: 
\begin{equation}
	\begin{array}{rcccl}
		{\cal L}_{\textrm{pure sugra}} =& \underbrace{{\cal L}_{\textrm{kin}}(g_{\mu\nu})} &+& \underbrace{{\cal L}_{\textrm{kin}}(\psi_\mu)} &+ {\cal L}_{\textrm{int}}(g_{\mu\nu},\psi_\mu), \\
		& \mbox{\footnotesize $  \frac{M_P^2}{2} \sqrt{-g} R$} && \mbox{\footnotesize $-\frac12 \overline \psi_\mu \gamma^{\mu\nu\rho} \partial_\nu \psi_\rho |_{\textrm{cov}}$} 
	\end{array}
	\label{Lsugra} 
\end{equation}
using 
\begin{equation}
	\delta g_{\mu\nu} \simeq \frac{1}{M_P} \overline \epsilon \gamma_{(\mu} \psi_{\nu)}|_{\textrm{cov}}, \label{deltagflat} 
\end{equation}
and 
\begin{equation}
	\delta \psi_\mu \simeq M_P \partial_\mu \epsilon|_{\textrm{cov}}, \label{deltapsi} 
\end{equation}
where \emph{cov} stands for a proper spacetime covariantization, and ${\cal L}_{\textrm{int}}$ denotes possible interaction terms that are not contained in the covariantizations of the kinetic terms (e.g., four-Fermion terms). 
This spacetime covariantization is done in the vierbein formalism (cf. Appendix \ref{subsec:Vierbein} and \ref{subsec:Spinorscurved}) and leads to the expressions (\ref{LSugraD4N1}), (\ref{varpsi}) and (\ref{susyvielbein}). Moreover, as we will see in Section \ref{cha:4DN1}, also the additional interaction terms not related to spacetime covariantization can elegantly be absorbed into the covariantized kinetic terms by working with covariant derivatives with non-trivial torsion. Before we come to this, however, let us briefly pause and take a quick look at some basic aspects of the gravitino field.

\subsection{Some remarks on the gravitino field and gravitino multiplets}
A vector-spinor field, $\psi_{\mu\alpha}(x)$, a priori has 16 degrees of freedom. 
The action for a free vector-spinor field in Minkowski spacetime is the Rarita--Schwinger action \cite{Rarita:1941mf}
\begin{equation}
	{\cal L}_{3/2} = - \frac12 \overline{\psi}_{\mu}\gamma^{\mu\nu \rho}\partial_\nu \psi_{\rho}  + \frac12 m_{3/2} \overline{\psi}_{\mu} \gamma^{\mu \nu} \psi_{\nu},
    \label{gravmass0}
\end{equation}
where $m_{3/2}$ is the physical mass of the corresponding particle in Minkowski spacetime. A careful analysis of the equations of motion shows that $\psi_{\mu\alpha}$ only propagates two physical degrees of freedom in the massless case, corresponding to states of helicity $\pm 3/2$, and four physical degrees of freedom in the massive case, corresponding to the polarization states of a massive spin-3/2 particle. 
The other degrees of freedom are either off-shell or auxiliary, as required to write a Lorentz-invariant action \cite{Fierz:1939ix}.

In the massless case, the Rarita--Schwinger action only consists of the kinetic term and is thus invariant under 
the gauge symmetry
\begin{equation}
	\delta \psi_\mu = \partial_\mu \Lambda,
    \label{psigaugeinv}
\end{equation}
where $\Lambda(x)$ is an arbitrary Majorana spinor. Just as for vector gauge bosons, this gauge invariance is necessary to eliminate longitudinal polarization states and has to be preserved by interactions so as to respect unitarity. In the context of supergravity, where $\psi_{\mu\alpha}(x)$ is the gravitino, the gauge symmetry (\ref{psigaugeinv}) is simply local supersymmetry.
 
While we encountered the gravitino as the superpartner of the helicity $\pm 2$ graviton, one might also wonder whether it would be possible to write down sensible field theories where the superpartner of a helicity $\pm 3/2$ particle is instead a helicity $\pm 1$ particle described by a vector field, $A_{\mu}(x)$. Such a multiplet is referred to as a \emph{gravitino multiplet}, and a globally supersymmetric free field theory for this multiplet indeed exists. As soon as one tries to introduce ineractions, however, the gauge invariance (\ref{psigaugeinv}) must be promoted to an additional local supersymmetry, and one arrives at a theory with several local supersymmetries, i.e. extended supergravity\footnote{Once more, consistent interacting theories including the gravitino may be obtained by allowing for higher spin gields and couplings between an infinite number of fields \cite{Vasiliev}, but this is not of interest for our discussion.}.

\section{Minimal supergravity in four dimensions} \label{cha:4DN1} 

\bigskip
Having clarified that local supersymmetry requires the coupling to gravity, we now want to show how to write down the minimal supersymmetric model, which involves only the gravity multiplet.
This represents the minimal supersymmetric extension of General Relativity, and provides a sufficient setup to investigate and illustrate a number of general features of supergravity theories, which remain valid in the presence of additional matter mutiplets, additional supersymmetries or more general  spacetime dimensions.

We will also show explicitly the calculations needed to prove supersymmetry invariance as they clarify the origin and meaning of a series of structures common to all supergravity theories.
Once more, additional details and extensions can be found in \cite{DallAgata:2021uvl}.

\subsection{The minimal action\protect\footnote{This section is reprinted from  \cite{DallAgata:2021uvl} © 2021 Springer-Verlag GmbH Germany, part of Springer Nature. Reproduced with permissions. All rights reserved.}}

In the following we will often use the language of differential forms in order to simplify calculations.
For instance, the action we want to supersymmetrize must contain the Einstein--Hilbert $S_{EH}$ and Rarita--Schwinger $S_{RS}$ actions,
\begin{equation}
	S = \int d^4 x \left({\cal L}_{EH}+{\cal L}_{RS}\right) = \int d^4x\,e\, \left(\frac{M_P^2}{2}\, R - \frac{1}{2} \overline{\psi}_{\mu }\gamma^{\mu\nu\rho}D_\nu \psi_{\rho }\right),
	\label{LSugraD4N1} 
\end{equation}
which, in the language of differential forms become 
\begin{equation}\label{Lsugraform}
	S = S_{EH}+S_{RS} = \frac{M_P^2}{4}\int R^{ab}\wedge e^c \wedge e^d \epsilon_{abcd} + \frac{i}{2}\int e^a \wedge \overline{\psi} \wedge \gamma_5 \gamma_a D\psi.
\end{equation}
Here, the indices $a,b,c,\ldots$ are local Lorentz indices referring to orthonormal frames, $e^a=e_{\mu}^{a}dx^{\mu}$, with constant epsilon tensor, $\epsilon_{0123}=1$, and $R^{ab}$ and $\psi$ denote, respectively, the curvature two-form (\ref{Riemannform})  of the spin connection and the gravitino one-form $\psi\equiv \psi_{\mu}dx^{\mu}$. $D$ denotes the Lorentz covariant derivative (\ref{covariantDspinor}). For further details on the formalism and our conventions, the reader is referred to the Appendix.  

Since we are coupling the spin $\nicefrac32$ field $\psi_{\mu}$ to gravity, the covariant derivative in its kinetic term should a priori be the full covariant derivative, $\nabla$, and not just the Lorentz-covariant derivative, $D$, we have used in the above expressions.
The full covariant derivative $\nabla$ contains both the Levi--Civita connection, $\Gamma$, coupling to the vector index $\mu$ of the gravitino, as well as the spin connection, $\omega$, coupling to the (suppressed) spinor index.
However, even if we had used $\nabla$, it would appear in the action only in anti-symmetrized form,\footnote{It is important here, that $\Gamma$ really denotes the torsion-free Levi--Civita connection.  As we will see later, it is useful to include a torsion piece bilinear in the gravitini in the \emph{spin connection} (but \emph{not} in the connection $\Gamma$, which should stay torsion-free). The connections defined by $\Gamma$ and $\omega$ are then no longer equivalent connections. } 
\begin{equation}
\nabla_{[\nu} \psi_{\rho]}  = \partial_{[\nu} \psi_{\rho]} + \frac14 \omega_{[\nu}^{ab} \gamma_{ab} \psi_{\rho]} - \Gamma_{[\nu\rho]}^\sigma \psi_{\sigma }, 
\end{equation}
and the last term is identically zero so that $\nabla_{[\nu} \psi_{\rho]}=  D_{[\nu}\psi_{\rho]}$, and we can indeed use the Lorentz-covariant derivative $D$ in the kinetic term of the gravitino.
In fact, the Levi--Civita connection in terms of Christoffel symbols will never really appear in the following. 

We now discuss the invariance under supersymmetry of (\ref{LSugraD4N1}).
We start by making one simple assumption that is motivated by our previous discussion on the gravitino being the gauge field of supersymmetry.
This means that the gravitino transformation rule should be proportional to the (covariant) derivative of the supersymmetry parameter
\begin{equation}
	\delta_\epsilon \psi_{\mu } = M_P D_\mu \epsilon \equiv M_P\left( \partial_\mu  + \frac14 \omega_\mu^{ab}\gamma_{ab} \right)\epsilon,
	\label{varpsi} 
\end{equation}
with the conjugate field satisfying
\begin{equation}
	\delta_{\epsilon} \overline{\psi}_{\mu } = M_P\left( \partial_\mu \overline{\epsilon} - \frac14\, \overline{\epsilon} \, \gamma_{ab}\, \omega_\mu^{ab} \right) \equiv M_P\,  \overline{D_\mu \epsilon}. 
\end{equation}

Having specified only the gravitino supersymmetry transformation so far, the next thing we would like to obtain is the transformation rule of the vierbein. 
We could simply make an educated guess in line with our
considerations leading to eq.~(\ref{deltag}), but let us try to actually derive the vierbein transformation law from what we already have.
{}From the variation of $S_{EH}$ we see that the only contribution with $\delta e^a$ comes multiplied by the curvature $R^{ab}$. 
We therefore try to single out from $\delta S_{RS}$ all possible terms that give the same type of contributions proportional to the curvature of the spin connection.
Supersymmetry invariance will then determine $\delta e^a$, and we will then check the invariance of the full action. 

The variation of the gravitini in the Rarita--Schwinger Lagrangian gives
\begin{equation}
\begin{array}{rcl}
	\delta {\cal L}_{RS} &=& \displaystyle- \frac{e}{2} \,\overline{\psi}_{\mu } \,\gamma^{\mu \nu \rho}\,  D_{\nu} \delta \psi_{\rho } - \frac{e}{2} \overline{\delta \psi_\mu}\,  \gamma^{\mu \nu \rho} \, D_{\nu} \psi_{\rho} + \ldots \\[3mm]
&=&\displaystyle	- \frac{e}{2} \,\overline{\psi}_{\mu } \,\gamma^{\mu \nu \rho}\,  D_{\nu} \delta \psi_{\rho } - \frac{e}{2} \overline{D_\nu \psi_\rho}\,  \gamma^{\mu \nu \rho} \, \delta \psi_{\mu } + \ldots \, , 
\end{array}\label{step1} 
\end{equation}
where we used the identity $\overline{\chi}\gamma^{\mu\nu\rho}\lambda=\overline{\lambda}\gamma^{\mu\nu\rho}\chi$ for anticommuting Majorana spinors, and the dots refer to the variations of the vierbein, $\delta e_\mu^a$, and the spin connection, $\omega_\mu^{ab}$, which we do not consider for now because they give terms that are not of the form we need.
Inserting (\ref{varpsi}) in (\ref{step1}) we obtain 
\begin{equation}
	\delta {\cal L}_{RS} = - M_P\, \frac{e}{2} \, \overline{\psi}_{\mu } \gamma^{\mu \nu \rho} D_{\nu} D_\rho \epsilon - M_P\, \frac{e}{2}\, \overline{D_\nu \psi_\rho} \gamma^{\mu \nu \rho} \,D_\mu \epsilon + \ldots \label{step2} 
\end{equation}
Integrating the last term by parts we can replace it by 
\begin{equation}
	- \partial_\mu \left(\frac{e}{2} M_P \, \overline{D_\nu \psi_\rho} \gamma^{\mu\nu\rho}\epsilon\right) + \frac{e}{2} \, M_P
\,  \overline{D_\mu D_\nu \psi_\rho} \gamma^{\mu\nu\rho}\epsilon \label{altrostep}
\end{equation}
plus terms involving derivatives of the vielbeine, $D_\mu e_\nu^a$, which we also neglect in this first step, because they will not give contributions proportional to the curvature $R^{ab}$. 
The equivalence of (\ref{altrostep}) to the last term in (\ref{step2}) can easily be checked either by recalling that the $\gamma$-matrices are covariantly constant in the sense that
\begin{equation}
	D_\mu \gamma^a = \partial_\mu \gamma^a + \omega_\mu{}^a{}_b \gamma^b + \frac14 \omega_\mu^{bc} [\gamma_{bc},\gamma^a] = 0, \label{relationDgamma} 
\end{equation}
or that $D_\mu(\hbox{scalar}) = \partial_\mu (\hbox{scalar})$. 
{}From the definition of the covariant derivative acting on fermions, we find 
\begin{equation}
	[D_\mu, D_\nu] = \frac14 R_{\mu\nu}{}^{ab} \gamma_{ab}, \label{integrD} 
\end{equation}
and therefore, using $\overline{\gamma_{ab}\psi_\rho}=-\overline{\psi_\rho}\gamma_{ab}$,
\begin{equation}
	\overline{D_{[\mu} D_\nu\psi_{\rho]}} = -\frac18 R_{[\mu\nu}{}^{ab} \overline{\psi}_{\rho]}\gamma_{ab}. 
	\label{integrDpsi} 
\end{equation}
Hence the variation of the Rarita--Schwinger term becomes 
\begin{eqnarray}
	\delta {\cal L}_{RS} &=& - \frac{e}{16} \,M_P \,\overline \psi_{\mu } \gamma^{\mu\nu\rho}\gamma_{ab}\epsilon \,  R_{\nu\rho}{}^{ab} - \frac{e}{16} \,M_P\, \overline{\psi}_{\rho }\gamma_{ab}\gamma^{\mu\nu\rho}\epsilon R_{\mu\nu}{}^{ab} 
	+ \ldots \nonumber \\[2mm]
	&=& -\frac{e}{16}\,M_P\, \overline{\psi}_{\mu }\left\{\gamma^{\mu \nu \rho}, \gamma_{ab}\right\} \epsilon R_{\nu \rho}{}^{ab}  
	+ \ldots \nonumber \\[2mm]
	&=& - \frac{e}{2} \,M_P\, \overline\psi_{\mu }\gamma^\nu \epsilon \left(R_\nu{}^\mu - \frac12\,\delta_\nu^\mu \,R\right) 
	+ \ldots  \\[2mm]
	&=& - \frac{e}{2} \,M_P\, \overline\psi_{\mu }\gamma^\nu \epsilon G_\nu{}^\mu + \ldots, \nonumber 
\end{eqnarray}
where we introduced the Einstein tensor $G_{\mu\nu}$ and used the identity 
		 $\{\gamma_{\mu\nu\rho}, \gamma^{ab}\} = -12 \, \gamma_{[\mu} e_{\nu}^a e_{\rho]}^b$. The dots contain all terms that do not multiply the curvature of the spin connection. 
As expected, this can be compensated by 
\begin{equation}
	\frac{\delta {\cal L}_{EH} }{\delta e_\mu^a} \delta e_\mu^a = -M_P^2\, e\, e_a^\nu\, G_{\nu}{}^\mu\, \delta e_\mu^a, \label{compens} 
\end{equation}
which is also proportional to the same combination of the curvature, provided we define the variation of the vierbein as 
\begin{equation}
	\delta e^a_\mu = \frac{1}{2 M_P}\overline \epsilon\gamma^a \psi_{\mu }
	\label{susyvielbein} 
\end{equation}
(recall that $\overline \epsilon \gamma^a \psi_{\mu } = - \overline \psi_{\mu } \gamma^a \epsilon$). 
Note that, proceeding in this way, we did not simply guess the variation of the vielbein from our considerations in the previous section, but instead really \emph{derived} it. 
On the other hand, we see immediately that (\ref{susyvielbein}) is indeed consistent with (\ref{deltag}).

In order to complete the proof of the invariance of the action (\ref{LSugraD4N1}), we still need to discuss the following variations:
\begin{itemize}
	\item[i)] $\displaystyle \frac{\delta {\cal L}_{EH}}{\delta \omega^{ab}_\mu} \delta_\epsilon \omega^{ab}_\mu$;
	\item[ii)] $\displaystyle \frac{\delta {\cal L}_{RS}}{\delta \omega^{ab}_\mu} \delta_\epsilon \omega^{ab}_\mu$; \phantom{$\displaystyle \stackrel{\int}{\int}$}
	\item[iii)] $\displaystyle \frac{\delta {\cal L}_{RS}}{\delta e_\mu^a} \delta_\epsilon e_\mu^a$; \phantom{$\displaystyle \stackrel{\int}{\int}$}
	\item[iv)] Terms involving $De^a$ from the partial integration in $\displaystyle \frac{\delta {\cal L}_{RS}}{\delta \psi_\mu} \delta_\epsilon \psi_\mu$. \phantom{$\displaystyle \stackrel{\int}{\int}$}
\end{itemize}
We also need to understand and specify $\delta_\epsilon \omega^{ab}$.
As we will see, the variation of the spin connection will depend on the formalism (first, second or 1.5 order) used to prove the invariance of the action.

To do this calculation, we go back to the form expression (\ref{Lsugraform}). 
The variation of the action is then
\begin{equation}
	\begin{array}{rcl}
		\delta S &=& \displaystyle \underbrace{\frac{M_P^2}{4} \,D \delta \omega^{ab} \wedge e^c \wedge e^d \epsilon_{abcd}}_{B1} + \underbrace{\frac{M_P^2}{2}\, R^{ab} \wedge \delta e^c \wedge e^d \epsilon_{abcd}}_{A1} \\[10mm]
		&& \displaystyle+ \underbrace{\frac{i}{2}\, \delta e^a \wedge \overline \psi \wedge \gamma_5 \gamma_a D\psi}_{B2} + \underbrace{\frac{i}{2}\, e^a \wedge \overline{\delta \psi} \wedge \gamma_5 \gamma_a D \psi}_{A2} \\[10mm]
		&&\displaystyle \underbrace{-\frac{i}{8} \, e^a \wedge \overline \psi \wedge \gamma_5 \gamma_a \gamma_{cd} \psi \wedge \delta \omega^{cd}}_{B3} + \underbrace{\frac{i}{2}\, e^a \wedge \overline \psi \wedge \gamma_5 \gamma_a D \delta \psi}_{A3} ,
	\end{array}
	\label{variationN1} 
\end{equation}
where the first line is the variation of $S_{EH}$, and the last two lines come from varying $S_{RS}$, with $B3$ being due to the variation of the spin connection inside $D$.

We know from previous computations that the term $A1$, coming from $\delta {\cal L}_{EH}/\delta e^a$, and the terms involving $D^2 \epsilon$, $D^2 \psi$ coming from  $\delta {\cal L}_{RS}/\delta \psi$ cancel. 
In detail, the $D^2 \epsilon$-term is $A3$, where one uses the explicit expression for $\delta \psi$, and the $D^2 \psi$-term can be extracted from $A2$ using the same steps that also led to (\ref{susyvielbein}).  
To do so, we switch the 2-form $D\psi$ and the 1-form $\delta \psi$, using (\ref{trasporules}), so that 
\begin{equation}
A2\equiv	 \frac{i}{2}\, e^a \wedge \overline{\delta \psi} \wedge \gamma_5 \gamma_a D \psi =   \frac{i}{2}\, e^a \wedge \overline{D \psi} \wedge \gamma_5 \gamma_a \delta \psi = 
\frac{i}{2} M_P\, e^a \wedge \overline{D \psi} \wedge \gamma_5 \gamma_a D \epsilon
	\label{expr1} 
\end{equation}
and, integrating again by parts, 
\begin{equation}
	\begin{array}{l}
		\displaystyle A2 =- M_P\, d \left(\frac{i}{2}e^a \wedge \overline{D \psi} \gamma_5 \gamma_a \epsilon\right) + \underbrace{\frac{i}{2}\, M_P\, D e^a \wedge \overline{D \psi}  \gamma_5 \gamma_a \epsilon}_{A2^{\prime\prime}} \\[2mm]
\displaystyle	\underbrace{- \frac{i}{2} M_P\, e^a \wedge \overline{D D\psi}  \gamma_5 \gamma_a \epsilon}_{A2^{\prime}} . 
	\end{array}
	\label{expr2} 
\end{equation}
The term $A2^{\prime}$ then cancels $A1$ and $A3$ as before, and we are left with $\delta S=B1+B2+B3+A2^{\prime\prime} $ plus boundary terms.
 
To proceed further, we integrate by parts the term $B1$ and get 
\begin{equation}
	B1= \frac{M_P^2}{2} \delta \omega^{ab} \wedge D e^c \wedge e^d \epsilon_{abcd} + d \left(\frac{M_P^2}{4} \delta \omega^{ab} \wedge e^c \wedge e^d \epsilon_{abcd}\right). 
\nonumber	
\end{equation}
In order to write $B3$ in a very similar form, we can use properties of the $\gamma$-matrices
\begin{equation}
\overline{\psi}\wedge \gamma_5 \gamma_a \gamma_{cd} \psi = \overline{\psi}\wedge \gamma_5 (\gamma_{acd} +\eta_{ac}\gamma_d-\eta_{ad}\gamma_c) \psi = -i\, \overline{\psi}\wedge \gamma^e\psi \epsilon_{acde}
\end{equation} 
so that, after some relabelling and reordering,
\begin{equation}
B3= -\frac{1}{8}\delta \omega^{ab}\wedge \overline{\psi}\wedge \gamma^{c}\psi\wedge e^d\epsilon_{abcd}.
\end{equation}
Discarding boundary terms and inserting also (\ref{susyvielbein}) in B2, we then have
\begin{equation}
	\begin{array}{rcl}
		\delta S &=& \displaystyle \frac{M_P^2}{2} \delta \omega^{ab} \wedge \left(De^c- \frac{1}{4M_P^2}\overline \psi\wedge \gamma^c \psi\right) \wedge e^d \epsilon_{abcd} \\[4mm]
		&& \displaystyle +\frac{i}{4 M_P} \,\left(\overline \epsilon \gamma^a \psi \right) \wedge (\overline{\psi} \wedge\gamma_5 \gamma_a D\psi)
		+\frac{i}{2} M_P \, D e^a \wedge \overline{D \psi} \gamma_5 \gamma_a \epsilon. 
	\end{array}
	\label{variation2} 
\end{equation}
This expression can be simplified by rewriting the second line so that the torsion piece $ \left(De^a- \frac{1}{4M_P^2}\overline \psi\wedge \gamma^a \psi\right)$ can also be factored out.
While the last term of (\ref{variation2}) obviously contains a derivative of the vielbein, the other term needs a reshuffling of the gravitini in order to produce the right bilinear without derivatives.
We can achieve this by using the Fierz identity 
\begin{equation}
	\psi  \,\wedge\, \overline{\psi} = \frac14 \, \left(\overline{\psi} \wedge \gamma^a \psi\right)\; \gamma_a - \frac18\,\left( \overline{\psi}\wedge\gamma^{ab} \psi\right) \; \gamma_{ab}
\end{equation}
and the fact that $\gamma^c \gamma^{ab} \gamma_c = 0$ and $\gamma^c \gamma^a \gamma_c = -2 \gamma^a$:
\begin{equation}
\frac{i}{4 M_P} \,\left(\overline \epsilon \gamma^a \psi \right) \wedge (\overline{\psi} \wedge\gamma_5 \gamma_a D\psi)= -\frac{i}{8M_P}(\overline{\psi}\wedge\gamma_{a}\psi)\wedge (\overline{D\psi}\gamma_{5}\gamma^{a}\epsilon).
\end{equation}
Altogether, $\delta S$ can then  be written as
\begin{eqnarray}
\hspace{-7mm}	\delta S &=& \frac{M_P}{2} \left(D e^a - \frac{1}{4 M_P^2} \overline\psi \wedge \gamma^a \psi\right) \wedge \left[i\overline{D\psi}\gamma_{5}\gamma_{a} \epsilon 
	+ M_P \delta \omega^{bc} \wedge e^d \epsilon_{abcd} \right]\nonumber\\[4mm]
&=& \frac{M_P}{2} \left(D e^a - \frac{1}{4 M_P^2} \overline\psi \wedge \gamma^a \psi\right) \wedge \left[-\frac{1}{6}\overline{D\psi}\gamma^{bcd} \epsilon 
	+ M_P \delta \omega^{bc} \wedge e^d \right] \epsilon_{abcd} ,\phantom{D}	\label{variationfinal}
\end{eqnarray}
where we have used $\gamma_{5}\gamma_{a}=(i/6) \epsilon_{abcd}\gamma^{bcd}$.
At this point the variation of the spin connection assumes a primary role and we can try to set (\ref{variationfinal}) to zero in various different ways.

\subsubsection{Second order formalism} 

\label{sub:second_order_formalism}

\index{second order formalism}

In this case, one imposes the so-called conventional constraint\index{conventional constraint},
\begin{equation}
	D e^a = \frac{1}{4 M_P^2} \overline \psi\wedge \gamma^a \psi, \label{def2order} 
\end{equation}
which determines the spin connection, $\omega_\mu^{ab}=\hat{\omega}_{\mu}^{ab}(e,\psi)$, as the solution to this equation. The spin connection is thus treated 
from the very beginning as a dependent field, whose supersymmetry variation follows from the supersymmetry variations 
 of $e_\mu^a$ and $\psi_\mu$ via the chain rule and the explicit functional dependence of $\hat{\omega}_{\mu}^{ab}(e,\psi)$.
By simple inspection of (\ref{variationfinal}), however, we see that (\ref{def2order}) already implies that $\delta_ \epsilon S = 0$, and we don't really need to know $\delta_\epsilon \omega_\mu^{ab}$.

Let us nevertheless use the torsion \index{torsion constraint}
\begin{equation}
	T^a = \frac{1}{4 M_P^2} \overline \psi \wedge \gamma^a \psi \label{torsion2order} 
\end{equation}
to solve $De^a = T^a$ for the spin connection\index{spin connection}\index{supercovariant connection}, which yields
\begin{equation}
	\hat{\omega}_\mu^{ab}(e, \psi) = \omega_\mu^{ab}(e) - \frac{1}{4 M_P^2}\left(\overline \psi^{[a} \gamma_\mu \psi^{b]} - \overline \psi_{\mu} \gamma^{[a}\psi^{b]} - \overline \psi^{[a} \gamma^{b]} \psi_{\mu}\right), \label{spinconn2order} 
\end{equation}
where $\omega_{\mu}^{ab}(e)$ is the torsion-free spin connection (\ref{spinconnection}), and the remaining piece is the contorsion tensor (\ref{contorsion}). 
This is used in the original approach of Ferrara, Freedman and Van Nieuwenhuizen in \cite{Freedman:1976xh}.
It is interesting to point out that the supersymmetry variation of this connection, inherited from the variations of $e_\mu^a$ and $\psi_\mu$, does not contain derivative terms, $\partial \epsilon$, of the supersymmetry parameter: 
\begin{equation}
	\delta \hat{\omega}_\mu^{ab} = \frac{1}{M_P}\, \overline \epsilon \gamma^\rho \left(D^\sigma \psi^\tau\right) \left(2 e_\rho^{[a} e_\tau^{b]} g_{\mu \sigma} -e_\tau^{[a} e_\sigma^{b]} g_{\mu \rho}\right). \nonumber
\end{equation} 
This is the reason why $\hat{\omega}_\mu^{ab}$ is often called  \emph{supercovariant}.

\subsubsection{First order formalism} 

\label{sub:first_order_formalism}

This is the approach followed by Deser and Zumino in their original paper \cite{Deser:1976eh}.\index{first order formalism}

Asking for the invariance of the action, $\delta S = 0$, via the vanishing of the term in square brackets in (\ref{variationfinal}) fixes the variation of the spin connection to 
\begin{equation}
	\delta \omega_\mu^{bc} = B_\mu^{bc} - \frac12 \, e_\mu^c B_e^{be} + \frac12 e_\mu^b B_e^{ce},
	\label{varspinconn} 
\end{equation}
with
\begin{equation}
	B_\mu^{bc} = \frac{i}{2 M_P} \overline \epsilon \gamma_\mu \gamma_5 D_{\rho} \psi_{\sigma} \epsilon^{\rho \sigma bc}.
    \label{Bmuab}
\end{equation}
This can be extracted using the same trick that is used in Appendix \ref{subsec:Vierbein} to derive the form (\ref{spinconnection}) of the torsion-free spin connection in terms of the vierbein from the torsion constraint.
It should be noted that (\ref{varspinconn}) is not the same as the variation derived using second order approach in the previous subsection. 
However, they become equivalent upon using the gravitino equations of motion.

\subsubsection{1.5-order formalism} 

\label{sub:1_5_order_formalism}

\index{1.5 order formalism}
In the 1.5 order formalism, one uses the fact that (\ref{def2order}) can be obtained as a field equation from varying the action with respect to $\omega_{\mu}^{ab}$ (as is obvious from the terms proportional to $\delta \omega^{bc}$ in (\ref{variationfinal})). Thus, when we determine the supersymmetry variation of the action and require $\omega_{\mu}^{ab}$ to be determined by  
\begin{equation}
D e^a = \frac{1}{4 M_P^2} \overline \psi\wedge \gamma^a \psi \Leftrightarrow \frac{\delta S}{\delta \omega_{\mu}^{ab}}=0, \nonumber	
\end{equation} 
we can immediately drop all terms proportional to $\delta \omega_{\mu}^{ab}$, as these are proportional to $\frac{\delta S}{\delta \omega_{\mu}^{ab}}$, which vanishes on-shell. Obviously, for the simple action we consider here, the only advantage over the second order formalism is that we would not have to keep track of the $\delta \omega_\mu^{ab}$ terms in (\ref{variationfinal}).
Just as in the second order formalism, the vanishing of the supersymmetry variation (\ref{variationfinal}) is thus obtained by using (\ref{def2order}), with the difference that (\ref{def2order}) is now not imposed by hand, but arises as a \emph{field equation} for the independent field $\omega_{\mu}^{ab}$. It is in this sense that the 1.5 order formalism combines elements from the first order formalism (the a priori independence of the field $\omega_{\mu}^{ab}$) and from the second order formalism (the use of (\ref{def2order}) for the cancellation of (\ref{variationfinal})).

It should be stressed that the 1.5 order trick of using on-shell field equations in the supersymmetry variation  can only be used for auxiliary fields such as $\omega_\mu^{ab}$.


%
%
%


\subsection{Gauging the Poincar\'e algebra\protect\footnote{This section is reprinted from  \cite{DallAgata:2021uvl} © 2021 Springer-Verlag GmbH Germany, part of Springer Nature. Reproduced with permissions. All rights reserved.}} 
\label{sec:appendix_gauging_the_poincar_e_algebra}

When introducing general relativity as well as supergravity, we discussed the possibility of considering the vierbein and the spin connection as independent quantities.
Do we have any conceptual reason behind this, in addition to the simplification of some computations? We will now see that an interesting perspective on gravity, which can help when dealing with supergravity, is that of considering gravity itself as a sort of a gauge theory where the gauge group is the Poincar\'e group \cite{Chamseddine77,MacDowell77,vanNieuwenhuizen:2004rh}.
This analogy will work only to a certain extent, but it will be very useful for understanding many specific new features that have to be introduced when one wants to promote supersymmetry to a local symmetry of nature.
In fact, supergravity \emph{is} the gauge theory of supersymmetry and therefore there must be a way to describe it as a theory where the gauge group is the Poincar\'e supergroup (or some other supergroup).
For the sake of simplicity in this subsection we set $M_P = 1$.  

Consider an ordinary gauge transformation $\delta_\epsilon = \epsilon^A T_A$, where $T_A$ are the gauge generators satisfying 
\begin{equation}
	[T_A,T_B] = f_{AB}{}^C T_C, \nonumber
\end{equation}
with structure constants $f_{AB}{}^C$. If this is a global symmetry of an action, it can be made local by introducing vector fields, $A_\mu^A$, for each symmetry so that the algebra 
\begin{equation}
	[\delta(\epsilon_1^A),\delta(\epsilon_2^B)] = \delta\left(\epsilon_2^B \epsilon_1^A f_{AB}{}^C\right), \label{gaugealgebra} 
\end{equation}
with symmetry parameters $\epsilon_{i}^{A}$ has a faithful realization on them, 
\begin{equation}
	\delta_\epsilon A_\mu^A = \partial_\mu \epsilon^A + \epsilon^C\, A^B_\mu\, f_{BC}{}^A, \label{vectorrealization} 
\end{equation}
and we can introduce covariant derivatives 
\begin{equation}
	D_\mu = \partial_\mu - A_\mu^A T_A \label{gaugecovariantderivative} 
\end{equation}
acting non-trivially on fields which transform in non-trivial representations of the gauge group.
The curvature, defined as 
\begin{equation}
	[D_\mu,D_\nu] = - F_{\mu\nu}^A T^A \qquad \Leftrightarrow \qquad F_{\mu\nu}^A \equiv 2
	\partial_{[\mu}A_{\nu]}^A + A_\mu^B A_\nu^C f_{BC}{}^A, \label{gaugecurvature} 
\end{equation}
transforms covariantly: 
\begin{equation}
	\delta_\epsilon F^A_{\mu\nu} = \epsilon^C F_{\mu\nu}^B f_{BC}{}^A.
\label{la2} 
\end{equation}

Let us now imagine that we want to make local the symmetries of the Poincar\'e group.
The usual procedure is to introduce gauge fields in correspondence with the generators of the algebra.
For the Poincar\'e algebra with generators $P_a$ and $M_{ab}$, this means introducing two gauge fields, $e_\mu^a$ and $\omega_\mu^{ab}$, so as to match the gauge generators, 
\begin{equation}
	A_\mu^A T_A = e_\mu^a P_a +\frac12\, \omega_\mu^{ab} M_{ab}.
\label{gaugePoinc} 
\end{equation}
Given the particular form of the Poincar\'{e} algebra, the  gauge curvatures of these vectors are precisely 
\begin{equation}
	T^a = d e^a + \omega^a{}_b \wedge e^b
\end{equation}
and 
\begin{equation}
	R^a{}_b = d \omega^a{}_b + \omega^a{}_c \wedge \omega^c{}_b,
\end{equation}
where the spin connection and the vierbein are so far independent fields.

This construction is perfectly legitimate.
However, it clearly leads to an ordinary gauge theory\footnote{Due to the Poincar\'{e} algebra being non-semisimple and non-compact, the standard kinetic terms of the gauge fields would not be positive definite so that this would actually not be a unitary theory.} and not to a gravity theory as we would like.
{}From the $T^a$ and $R^{ab}$ curvatures we could construct kinetic terms giving the propagation of independent degrees of freedom and discuss the resulting gauge theory, where the Poincar\'e group is realised on the vector fields as 
\begin{eqnarray}
	\delta_P \omega^{ab} & =& 0, \label{transl_omega}\\[2mm]
	\delta_P e^{a} & =& D \epsilon^a, \label{transl_e}\\[2mm]
	\delta_M \omega^{ab} & =& d \Lambda^{ab} - \omega^{a}{}_c \Lambda^{cb} - \Lambda^{ac}\omega_c{}^b, \label{rotation_omega}\\[2mm]
	\delta_M e^{a} & =& \Lambda^a{}_c e^c,
\label{rotation_e} 
\end{eqnarray}
with gauge parameters $\epsilon^a$ and $\Lambda^{ab}$.
If, on the other hand, we want to get only the metric degrees of freedom we have to 
impose a constraint between $\omega^{ab}$ and $e^a$.
The constraint that does this job is the \emph{conventional constraint}
\index{conventional constraint} or torsion constraint (see \cite{Kibble,vanNieuwenhuizen:2004rh} for details)÷,
\index{torsion constraint} 
\begin{equation}
	T^a = 0.
\label{convconst} 
\end{equation}
This constraint, however, is not invariant under (\ref{transl_omega})--(\ref{transl_e}):
\begin{equation}
	\delta_P T^a = \delta_P D e^a = D \delta_P e^a = DD \epsilon^a = -R^a{}_b \epsilon^b \neq 0.
\label{la3} 
\end{equation}
This means that if we impose the conventional constraint, translation symmetry is broken.
Moreover, it is also clear that now the spin connection $\omega^{ab}$ cannot be treated as independent of the vielbein anymore and hence the transformation (\ref{transl_omega}) will no longer be valid.
Indeed, since $\omega^{ab} = \omega^{ab}(e)$, the spin connection is not invariant under translations 
\begin{equation}
	\delta_P \omega^{ab} = \int d^4 x\, \frac{\delta \omega^{ab}}{\delta e^c} \delta_P e^c \neq 0. \nonumber
\end{equation}
The final outcome of this discussion is that, when the conventional constraint is imposed, the Poincar\'e gauge algebra is deformed and translational symmetry is replaced by a new invariance under diffeomorphisms.
This can be seen by considering the commutator of two translation generators on the vierbein: 
\begin{equation}
	[\delta_{P2},\delta_{P1}] e^a = - \delta_{P[2}(D \epsilon_{1]}^a) = - \delta_{P[2} \omega^{a}{}_c \epsilon_{1]}^c, \label{transl_commutator} 
\end{equation}
and now $\delta_P \omega^{ab} \neq 0$.
The resulting algebra then has a non-vanishing commutator 
\begin{equation}
[P,P] \neq 0, \nonumber	
\end{equation} 
as is appropriate for general coordinate transformations, which do not commute.
Actually one can check that using the constraint (\ref{convconst}) the translation generators on the vielbein take the form of general coordinate transformations.

The action constructed from the curvatures and the vierbein then is invariant with respect to local Lorentz transformations and diffeomorphisms.
The infinitesimal change of a function under a diffeomorphism is given by the Lie derivative 
\index{Lie derivative} ${{L}}_\epsilon$, and therefore the action is going to be invariant if (cf. Appendix \ref{subsec:Liederivative}) $$ {{L}}_\epsilon S = \int d (\imath_\epsilon {\cal L}) + \int \imath_\epsilon d {\cal L} = 0,$$ but the first term is a total derivative that can be discarded while the second is zero because $d{\cal L}$ has one degree more than the top-form.
Finally, in the construction of an action, we will not make use of a kinetic term of the form $R^{ab} \wedge \star R_{ab}$ because of the conventional constraint which makes it quartic in the derivatives.
The appropriate quadratic term is the Einstein--Hilbert action above.

The method we have outlined in this section can be easily extended to generic supergravity theories by extending the Poincar\'{e} algebra to the super Poincar\'{e} algebra by including fermionic generators and possibly other bosonic generators for the internal symmetries.
The power of this approach lies in the ease of guessing the transformation laws under the various symmetries, including supersymmetry.
This means that this approach can be used as a guide to derive and construct the Lagrangian and/or the equations of motion of systems respecting any symmetry group we would like to realize. 
Once again, we stress that one has to be careful with its application because of the constraints that will be needed to obtain a consistent gravity theory (invariant under diffeomorphisms).
Imposing these constraints will break the transformation rules that do not preserve them.

We end this discussion with a few remarks on the gauging of the \emph{super} Poincar\'{e} group.
We could gauge this algebra by adding new vector fields $\psi_{\mu A}$ for the fermionic generators $Q^A$.
{}From the algebra we then have 
\begin{equation}
	A^A_\mu T_A = e_\mu^a P_a + \frac12\,\omega_\mu^{ab} M_{ab} + \overline \psi_{\mu A} Q^A + \overline \psi_\mu^A Q_A, 
\end{equation}
and we can read the supersymmetry transformations by applying $$ \delta_\epsilon A_\mu^A = 
\partial_\mu \epsilon^A + \epsilon^C \,A^B_\mu \,f_{BC}{}^A.$$ 
For instance, for ${\cal N} = 1$ supergravity, we would get that the spin connection is invariant, 
\begin{equation}
	\delta_\epsilon \omega^{ab} = 0, \label{susy_spin} 
\end{equation}
because the Lorentz generator never appears on the right hand side of any commutator involving the supersymmetry generator.
However, just like for the bosonic case, we should impose a torsional constraint in order for the vierbein and spin connection not to be independent.
Doing so, we would fix the form of the spin connection as $\omega^{ab} = \omega^{ab}(e,\psi)$ and could check the new realization of the algebra on the fields.

One last interesting remark involves the definition of the gauge curvatures for the super-Poincar\'e algebra.
{}From the structure constants of the supersymmetry algebra, one can deduce a new definition for the curvatures, including the one of the translation generators, which is usually denoted as the 2-form ${\cal T}^a$.
Since the translation generators $P_a$ appear on the right hand side of the commutator of two supercharges the corresponding curvature definition is now
\begin{equation}
	{\cal T}^a = D e^a - \frac14\, \overline \psi^A \gamma^a \psi_A
\end{equation}
and involves a fermion bilinear.
This means that imposing the constraint ${\cal T}^a = 0$ results in a spin connection depending on the gravitino fields.  
Hence, supergravity is often referred to as a theory with non-trivial torsion for the spin connection, because ${\cal T}^a = 0$ implies $De^a \neq 0$.

\subsection{Adding a cosmological constant\protect\footnote{This section is reprinted from  \cite{DallAgata:2021uvl} © 2021 Springer-Verlag GmbH Germany, part of Springer Nature. Reproduced with permissions. All rights reserved.}} 

\label{sec:adding_a_cosmological_constant}

So far we considered the construction of a supergravity action around a Minkowski background, whose non-linear completion led to Einstein gravity coupled to a gravitino field without a cosmological constant.
In ordinary Einstein gravity, however, we can always add a cosmological constant\index{cosmological constant} $\Lambda$ to obtain the action
\begin{equation}
	S = M_P^2 \int d^4 x \sqrt{-g} \left(\frac{1}{2} R - \Lambda\right)
	\label{startact} 
\end{equation}
and find a maximally symmetric vacuum with \index{de Sitter space}\index{Anti-de Sitter space}
\begin{equation}
	R_{\mu\nu} = \Lambda g_{\mu\nu} \qquad 
	\begin{array}{lcl}
		\Lambda > 0 & \leftrightarrow & \hbox{de Sitter (dS)},\\[5mm]
		\Lambda < 0 & \leftrightarrow & \hbox{Anti-de Sitter (AdS)}. 
	\end{array}
	\label{AdSvac} 
\end{equation}

It is natural to ask whether these solutions and the corresponding actions can be supersymmetrized in a natural way.
In this section, we will focus on pure supergravity theories (without matter multiplets).
If we want to construct a supergravity action generalizing (\ref{startact}), we should be able to find a supergroup that contains the symmetry group of AdS and/or dS spacetime. 
We will now see that minimal supersymmetry constrains the closure of the algebra in a way that only one of the two options is consistent\footnote{There are consistent de Sitter superalgebras with extended supersymmetries, but they do not allow for positive weight representations and hence their realizations have the wrong sign in front of the kinetic terms of some of their fields \cite{Ferrara:1977sc,Pilch:1984aw}.}.

Before discussing the corresponding superalgebras, we note that the symmetry groups of both AdS and dS in $d$ dimensions, SO(1,$d$) and SO(2,$d-1$), respectively, can be embedded in  SO$(2,d)$.
For $d = 4$, the (A)dS algebra is described by 10 anti-Hermitian generators $M_{\underline{AB}}$ satisfying the commutator relations
\begin{equation}
	[M_{\underline{AB}}, M_{\underline{CD}}] = - 2\,\eta_{\underline{C[A}} M_{\underline{B]D}} +2\, \eta_{\underline{D[A}} M_{\underline{B]C}}, \label{comm} 
\end{equation}
with $\eta_{\underline{AB}} = {\rm diag}\{-+++-\}$ for AdS, and $\eta_{\underline{AB}} = {\rm diag}\{-++++\}$ for dS space. 
The explicit (A)dS algebra follows by identifying $M_{5a} ={\ell} \, P_a$, where we split $\underline{A}= \{a,5\}$, with $a,b,\ldots=0,1,2,3$, and $\ell$ is the radius of curvature of (A)dS:
\begin{equation}
	\begin{array}{l}
		[M_{ab},M_{cd}] = - 2\, \eta_{c[a} M_{b]d} +2\, \eta_{d[a} M_{b]c}, \\[2mm]
		[P_a,M_{bc}] = 2\,\eta_{a[b} P_{c]}, \\[2mm]
\displaystyle		[P_a,P_b]= \pm \frac{1}{\ell^2} \,M_{ab} ,
	\end{array}
	\label{AdSalgebra} 
\end{equation}
where the last commutator is equivalent to $[M_{5a},M_{5b}] = - \eta_{55} M_{ab}$ and $\eta_{55} =-1$ for AdS space and $\eta_{55}=+1$ for dS. Hence, the upper sign is for AdS and the lower one for dS spacetime. 
Clearly, when $\ell \to \infty$ the (A)dS curvature goes to zero and one gets back the Poincar\'e algebra. 

To construct the full superalgebra, one needs to specify also the commutators with the supercharges.
In particular, $[P_a, Q]$ \emph{cannot} be zero anymore, as it used to be in the super-Poincar\'e case, because we would no longer close the super Jacobi identities, as
\begin{equation}
	[\underbrace{[P_a,P_b]}_{\sim M_{ab}},Q]+[\underbrace{[P_a,Q]}_0,P_b] -[\underbrace{[P_b,Q]}_0,P_a]  =0. \label{Jacnotclose} 
\end{equation}
We therefore need to impose new commutator relations such that the momenta do not commute with the supercharges.
To respect Lorentz covariance and the graded algebra structure, the result of the commutator should be proportional to the supercharges and come with some gamma matrices.
In principle there are two possibilities that respect the Majorana condition on the supercharges
\begin{equation}
		[P_a, Q] \sim \gamma_a Q, \qquad {\rm or} \qquad [P_a, Q] \sim \gamma_a \gamma_5 Q.
\end{equation}
In the first case the coefficient multiplying the right hand side should be real, while in the second it should be imaginary.
If we use a chiral notation, the sign and the ambiguities can be reabsorbed in a single dimensionful complex coefficient $\tilde{g}$.
We stress this fact, because there are sometimes wrong statements in the literature about this.
Once we introduce the chiral notation, the new commutators are
\begin{equation}
	[P_a, Q_R] = -\frac{\tilde{g}}{2}\,  \gamma_a Q_L, \qquad [P_a,Q_L] = -\frac{\tilde{g}^*}2\, \gamma_a Q_R. \label{impcom} 
\end{equation}
Once we introduce these new non-trivial commutators, the super Jacobi identity can be satisfied, though only for the AdS case.
This is readily seen by explicitly computing the results of the various commutators:
\begin{equation}
	\begin{array}{rcl}
		0 &\stackrel{!}{=}& [P_a,[P_b,Q_L]] + [Q_L,[P_a,P_b]]-[P_b,[P_a,Q_L]] \\[2mm]
		&=& \displaystyle \frac{|g|^2}{4}\, \left(\gamma_b \gamma_a - \gamma_a \gamma_b\right)Q_L \pm \frac{1}{\ell^2} [Q_L, M_{ab}] \\[2mm]
		&=& \displaystyle- \frac{|g|^2}{2}\, \gamma_{ab} Q_L \pm \frac{1}{2\ell^2} \gamma_{ab} Q_L,
	\end{array}
	\label{JacdS} 
\end{equation}
where we used the commutation relations in (\ref{susyalgebra}), but with the anti-Hermitian generators $P_a = i\,{\cal P}_a$ and $M_{ab} = i\, {\cal M}_{ab}$.
It is now clear that \emph{only for the plus sign} we can get a solution: 
\begin{equation}
	|g|^2 = \frac{1}{\ell^2}. 
	\label{solJac} 
\end{equation}
Hence, only for AdS we can write a consistent supersymmetric completion with a single supercharge.
We finally note that the closure of the super-Jacobi identites requires that another commutator gets modified, namely
\begin{equation}
		\{Q_L, \overline{Q}_L\} = \tilde{g}^* \gamma^{ab} M_{ab}.
		\label{QQcomm}
\end{equation}

The constraint imposing that the superalgebra can be defined only for the AdS supergroup and not for dS implies a very important fact: a positive cosmological constant will \emph{always} break supersymmetry, while a negative cosmological constant may be compatible with supersymmetry.

Although matter couplings or extended supersymmetries may allow for de Sitter vacua in a supersymmetric theory, the vacuum itself will always break supersymmetry.

Once supersymmetry is broken, one could describe this phase of the theory by using non-linear realizations, as it is customary for any other symmetry whose linear action is broken.
This has been the subject of intense scrutiny (see for instance \cite{Antoniadis:2014oya,Bergshoeff:2015tra,DallAgata:2016syy,Cribiori:2017ngp}) and one can indeed write actions for theories with dS vacua where supersymmetry is non-linearly realized.
Since an effective discussion of this topic requires some additional technical introduction to superfields in supergravity, we will not deal with it here, but refer the reader to the literature on the subject, such as \cite{Wess:1992cp}.

\bigskip

\subsubsection{Construction of the action} 

\label{sub:construction_of_the_action}

Now that we established that the anti de Sitter group can be consistently extended to a supergroup, we would like to realize it in terms of a supersymmetric action that includes a negative cosmological constant.
We will proceed in a fashion similar to what has been done in the flat case, starting from the supersymmetry transformation of the gravitino and then trying to close the action of the supersymmetry transformation on the free Lagrangian for the gravity multiplet, possibly introducing interaction terms.
We therefore need to fix first the supersymmetry transformation rule of the gravitino.
Since the algebra has been modified with respect to the case without cosmological constant, we expect that also the supersymmetry transformations get modified accordingly.

As sketched in Section \ref{sec:appendix_gauging_the_poincar_e_algebra}, we can generically deduce the supersymmetry transformation properties of the various fields from the structure constants of the underlying superalgebra. For supergravity without cosmological constant, one would do this by viewing supergravity as a gauge theory of the super-Poincar\'e group, but with some constraints needed to relate the vierbein and the spin-connection degrees of freedom. 

By using this trick, we can now deduce the supersymmetry transformation of the gravitino in the presence of a negative cosmological constant by looking at the structure constants of the $AdS$ superalgebra coming from commutators which have a supersymmetry generator on the right hand side.
This inspection shows that a new term in the supersymmetry transformation of the gravitino should appear because of the non-zero commutator (\ref{impcom}) between the translation generators and supersymmetry generators.
If one interprets the spin connection term in the Lorentz covariant derivative in the original gravitino transformation (\ref{varpsi}) as due to the non-vanishing commutator of $M_{ab}$ with $Q$, the new non-vanishing commutator (\ref{impcom}) between $P_a$ and $Q$ should then analogously lead to an additional contribution to the gravitino transformation so as to make the transformation covariant with respect to the full AdS isometry group. 
In $\delta\psi_{\mu L}$, this additional contribution should be of the form given in the first equation of (\ref{impcom}) contracted with the gauge field of the translation generator $P_a$, i.e.~with the vierbein $e_\mu^a$. 
We therefore should have
\begin{equation}
	\delta \psi_{\mu L} = M_P D_\mu \epsilon_L - \frac{g}{2} \, M_P^2\, \gamma_\mu \epsilon_R,
	\label{susygravcc} 
\end{equation}
where now we have a dimensionless constant $g\in {\mathbb C}$, because of the introduction of the dimensionful $M_P$ and $M_P^2$ factors.
Clearly, this has to go together with the conjugate relation: 
\begin{equation}
	\delta \psi_{\mu R} = M_P D_\mu \epsilon_R - \frac{g^*}{2} \, M_P^2\, \gamma_\mu  \epsilon_L. 
	\label{susygravccconjugate} 
\end{equation}
Once again we stress that $g$ here can be any complex number, because there are sometimes wrong statements in the literature about this.

To construct the action, we start from the action (\ref{LSugraD4N1}) with the vierbein transformation rule 
\begin{equation}
	\delta e^a_\mu = \frac{1}{2M_P}\overline \epsilon_L \gamma^a \psi_{\mu R} + h.c., \nonumber
\end{equation} 
and (\ref{susygravcc}) for the gravitino
\begin{equation}
	\delta \psi_{\mu L} = M_P D_\mu \epsilon_L - \frac{g}{2} \, M_P^2\, \gamma_\mu \epsilon_R. 
	\nonumber
\end{equation}
The reason we start from the action without the cosmological constant, rather than adding explicitly the cosmological constant among the bosonic terms right from the beginning, is that it will automatically be enforced by supersymmetry in an iterative procedure at higher order in $g$, as will become clear momentarily.

The gravitino relation differs from the one in (\ref{varpsi}) by a \emph{shift term}\index{shift term} proportional to the constant $g$. 
Clearly this shift breaks the supersymmetry of the original action (\ref{LSugraD4N1}) and we need to restore it by adding additional terms to it.
In the following, we will establish again the invariance under supersymmetry of a modified action.
To our knowledge this was first done in \cite{Townsend:1977qa}.

The first supersymmetry-breaking effect of the  shift term (proportional to $g$) is that of generating new terms in the variation of the Rarita--Schwinger part of the Lagrangian.
To compute these terms, we use the supersymmetry variation of the conjugate gravitino, which, in form notation, reads 
\begin{equation}
	\delta \overline \psi_R = M_P \overline{D \epsilon}_R + \frac{g^*}{2} \, M_P^2\,\overline \epsilon_L \gamma_a e^a, 
	\label{varpsiconj} 
\end{equation}
as one may easily verify.
Denoting by $\delta_g$ the variations due to the $\mathcal{O}(g)$ shift term in the gravitino transformation law,
the uncancelled variation of ${\cal L}_{RS}$ under supersymmetry is then
\begin{equation}
	\begin{array}{rcl}
		\delta_g {\cal L}_{RS} & = &\displaystyle \frac{i}{2} e^a \wedge \delta_g \overline \psi_R \wedge \gamma_5 \gamma_a D \psi_L + \frac{i}{2}\, e^a\wedge \overline \psi_L\wedge  \gamma_5 \gamma_a D \delta_g \psi_R + h.c. 
		\\[3mm]
		& =& \displaystyle \frac{i}{4} g^* \,M_P^2\, e^a \wedge e^b\wedge \overline \epsilon_L \gamma_b \gamma_5 \gamma_a D \psi_L \\[3mm]
		&& - \frac{i}{4} g^* M_P^2 \,e^a\wedge \overline \psi_L \wedge \gamma_5 \gamma_a D \left(e^b \gamma_b \epsilon_L\right) + h.c.\\[3mm]
		& = &\displaystyle \frac{i}{4} g^*\, M_P^2e^a \wedge e^b\wedge \overline \epsilon_L \gamma_5 \gamma_{ab} D \psi_L -\frac{i}{4} g^*\, M_P^2 \,e^a\wedge D e^b\wedge \overline \psi_L \gamma_5 \gamma_a \gamma_b \epsilon_L\\[3mm]
		&&\displaystyle - \frac{i}{4} g^*\, M_P^2 \,e^a \wedge e^b\wedge \overline \psi_L\wedge \gamma_5 \gamma_{ab} D\epsilon_L + h.c. 
	\end{array}
	\label{varRS} 
\end{equation}
Integrating by parts the first term in the last equality, we get 
\begin{eqnarray}
	\delta_g {\cal L}_{RS} &=& \frac{i}{2} g^*\, M_P^2\, e^a\wedge  D e^b\wedge  \left(\overline \epsilon_L \gamma_5 \gamma_{ab} \psi_L - \frac12 \overline \psi_L \gamma_5 \gamma_a \gamma_b \epsilon_L\right) \nonumber\\
	&& - \frac{i}{2} g^*\, M_P^2 e^a \wedge e^b \wedge\overline \psi_L \wedge \gamma_5 \gamma_{ab} D\epsilon_L + h.c. 
	\label{leftover} 
\end{eqnarray}
The first term, proportional to $De^a$, plays a similar role as in the case without a cosmological constant and will be discussed later after equation (\ref{deltaSS}).
Since the remaining terms are proportional to the derivative of the supersymmetry parameter, we can try and use the supersymmetry transformation rule of the gravitini, $\delta \psi_L = M_P D \epsilon_L + {\cal O}(g)$, to cancel them.
For this reason we add a mass-like term to the Lagrangian: \index{gravitino mass}
\begin{equation}
	{\cal L}_{{\cal M}_\psi} = \frac{i}{4} g^* \,M_P\, e^a\wedge e^b\wedge \overline \psi_L \wedge\gamma_5 \gamma_{ab} \psi_L + \frac{i}{4} g\, M_P\, e^a\wedge e^b\wedge \overline \psi_R\wedge \gamma_5 \gamma_{ab} \psi_R,
	\label{massterm} 
\end{equation}
so that the variation of $\psi$ in (\ref{massterm}) compensates for (\ref{leftover}) at order $g$.
We point out here the reduced Planck mass factors, so that the whole coefficient has mass dimension 1, as well as the overall factor $\nicefrac14$, which is due to the double variation required to match (\ref{leftover}).

While the introduction of (\ref{massterm}) allows the cancellation of the $D\varepsilon$ terms in (\ref{leftover}), it also gives rise to two further  new variations we have to take care of. 
One variation comes at order $g$ from the variation of the vierbeine in (\ref{massterm}). 
We will discuss its cancellation together with the cancellation of the $De^a$ terms in (\ref{leftover}) further below. The other new variation of (\ref{massterm}) is of order $g^2$ and arises when the order $g$ shift term in the gravitino transformation is used in the variation of the gravitini in (\ref{massterm}).
Indeed, the order $g^2$ variation of the gravitino mass term produces (suppressing the wedges)
\begin{equation}
	\delta_g {\cal L}_{{\cal M}_\psi} = -\frac{i}{8} |g|^2 M_P^3 e^a e^b \left(\overline \psi_L \gamma_5 \gamma_{ab} e^c \gamma_c \epsilon_R - \overline \epsilon_R e^c \gamma_c \gamma_5 \gamma_{ab}\psi_L\right) + h.c. 
	\label{varmassterm} 
\end{equation}
Putting together the gamma matrices and using the duality relation $\gamma_5 \gamma_{abc} = -i\, \epsilon_{abcd} \gamma^d$ we obtain 
\begin{equation}
	\begin{array}{rcl}
		\delta_g {\cal L}_{{\cal M}_\psi} &=&\displaystyle M_P^3 \,\frac{|g|^2}{4}\, e^a e^b e^c \epsilon_{abcd} \left(\overline \psi_L \gamma^d \epsilon_R + \overline \psi_R \gamma^d \epsilon_L\right) \\[2mm]
		&=&\displaystyle -M_P^4\, \frac{|g|^2}{2}\, e^a e^b e^c \epsilon_{abcd} \frac{\overline \epsilon_R \gamma^d \psi_L + \overline \epsilon_L \gamma^d \psi_R}{2 M_P}\\[2mm]
		&=&\displaystyle -M_P^4\, \frac{|g|^2}{2}\, e^a e^b e^c \epsilon_{abcd}\delta e^d\\[2mm]
		&=&\displaystyle - |g|^2 \frac{M_P^4}{8}\delta\left( e^a e^b e^c e^d\epsilon_{abcd}\right).\\[2mm]
	\end{array}
	\label{varmassterm2} 
\end{equation}

We further recall from  Appendix \ref{subsec:Vierbein} that $e^a e^b e^c e^d \epsilon_{abcd} = +4! \, d^4 x \, e$, and then realize that we need to add a single term of order $|g|^2$ to the Lagrangian to cancel (\ref{varmassterm2}):
\begin{equation}
	3 \int d^4 x \,e\, M_P^4 |g|^2 = -M_P^2 \int d^4 x\, e\, \Lambda. 
	\label{cc} 
\end{equation}
This is a \emph{cosmological constant term}. \index{cosmological constant}
Notice that there is \textbf{no choice of the sign} of this cosmological constant
\begin{equation}
	\Lambda = -3 \,M_P^2\, |g|^2 = - \frac{3}{\ell^2}<0. 
	\label{ccval} 
\end{equation}
This agrees with the discussion following from the supersymmetry algebra.

It is also extremely important to note that the variation of (\ref{cc}) does not generate terms of order $g^3$ and that therefore supersymmetry closes at order $g^2$.

The only variation left is the vierbein variation in the gravitino mass term together with the already mentioned first term in (\ref{leftover}).
Using 
\begin{equation}
4(\overline{\psi}_R\wedge\gamma_{ab}\psi_R)\wedge(\overline{\varepsilon}\gamma^b\psi) = 3(\overline{\psi}_R\gamma_{ab}\varepsilon_{R})\wedge(\overline{\psi}\wedge\gamma^b\psi)+(\overline{\psi}_R\varepsilon_R)\wedge(\overline{\psi}\wedge\gamma_a\psi),
\end{equation}
which one can derive from the Fierz identities (\ref{FierzRR}) - (\ref{cyclicid}), one obtains for all remaining uncancelled variations
\begin{eqnarray}
	&&\delta {\cal L} = \nonumber \\[2mm]  
	&&\frac{M_P}{2} \left(D e^a - \frac{1}{4 M_P^2} \overline\psi \wedge \gamma^a \psi\right) \wedge \left[i\, g^*\, M_P \, e^b \wedge \left(\overline \epsilon_L \gamma_{ba} \psi_L - \frac12 \overline \psi_L  \gamma_b \gamma_a \epsilon_L\right) \right. \nonumber\\[2mm]
	&&  - i\, g\, M_P\, e^b \wedge  \left(\overline \epsilon_R \gamma_{ba} \psi_R - \frac12 \overline \psi_R  \gamma_b \gamma_a \epsilon_R\right)  \label{deltaSS}  \\[2mm]
	&&\left.	+ \epsilon_{abcd}\left(	-\frac{1}{6}\overline{D\psi}\gamma^{bcd} \epsilon
	+ M_P \delta \omega^{bc} \wedge e^d \right) \right], \nonumber
\end{eqnarray}
where the last term proportional to $\varepsilon_{abcd}$ is the same as in the case without cosmological constant.

This completes the proof of the invariance of the action in any of the formalisms described above.
In the second order formalism this variation vanishes because of the torsion constraint.
In the first order formalism we deduce from this variation the expression for $\delta \omega^{ab}$ that makes it vanish.
Finally, in the 1.5 formalism, the equations of motion for the spin connection do not change and hence once again the full Lagrangian is invariant under supersymmetry.

Bringing the mass term (\ref{massterm})  to the standard form without differential forms, the final Lagrangian is therefore the following
\begin{eqnarray}
	{\cal L}&=& \frac{M_P^2}{2} e R-\frac{e}{2} \overline{\psi}_{\mu R} \gamma^{\mu\nu\rho}D_\nu \psi_{\rho L}-\frac{e}{2} \overline{\psi}_{\mu L} \gamma^{\mu\nu\rho}D_\nu \psi_{\rho R} \nonumber \\[2mm]
	&&- e\, M_P\, \frac{g}{2}\, \overline{\psi}_{\mu R} \gamma^{\mu\nu}\psi_{\nu R}- e \, M_P\, \frac{g^*}{2}\, \overline{\psi}_{\mu L} \gamma^{\mu\nu}\psi_{\nu L} 
	\label{SugraLambda}\\[2mm]
	&& + 3 \,e\,M_P^4\, |g|^2, \nonumber
\end{eqnarray}
where we should remember that in the second order formalism $\omega^{ab} = \omega^{ab}(e,\psi)$.
The final supersymmetry transformations are
\begin{eqnarray}
\delta e_\mu^a &=&\frac{1}{2M_P}\overline \epsilon_L \gamma^a \psi_{\mu R} + h.c.,\\[2mm]
\delta \psi_{\mu L} &= & M_P D_\mu \epsilon_L - \frac{g}{2} \, M_P^2\, \gamma_\mu \epsilon_R. 
\end{eqnarray}

Now that we completed the construction of a supersymmetric action for supergravity with a (negative) cosmological constant, we can make some comments.

First of all, we have seen from the construction that we have performed only minimal modifications.
After shifting the supersymmetry transformation of the gravitino field, we introduced the smallest set of terms needed to cancel ${\cal O}(g)$ and ${\cal O}(g^2)$ terms in the supersymmetry variations.
As already pointed out above, supersymmetry closes at order $|g|^2$. 
There is no need to introduce any term of order $g^3$ or more.

All the modifications can be summarized in three main pieces:
\begin{itemize}
	 \setlength\itemsep{1em}
	\item a shift in the fermionic supersymmetry rules at ${\cal O}(g)$; 
	\item a mass-like term of ${\cal O}(g)$ for the fermions; 
	\item a potential term at order ${\cal O}(g^2)$. 
\end{itemize}
Although these modifications have been forced by the presence of the cosmological constant in a pure gravity theory, one finds the pattern outlined above also in all gauged supergravity theories for models with extended supersymmetries.
In fact, in extended supergravities the appearance of non-abelian gauge groups is tied to the presence of a non-trivial scalar potential, which may act as an effective cosmological constant.
The result is that the gauging\index{gauging} procedure introduces the same three main modifications listed above, where the mass-like term for the fermions in the general case with scalar fields becomes a Yukawa-like coupling, and the cosmological constant term becomes a scalar potential.

It can also be seen that this scalar potential (in this case a pure cosmological constant) can be expressed as the square of the shifts of the supersymmetry variations of the fermionic fields: 
\begin{equation}
	V \,\overline \epsilon_L \gamma^a\epsilon_R = -3 M_P^4 |g|^2 \overline \epsilon_L \gamma^a \epsilon_R = -\frac32\, \delta_g \overline \psi_{\mu R}\gamma^a\delta_g \psi_L^\mu.
	\label{susyward0} 
\end{equation}
Note the minus sign in front of the squared gravitino shifts. 
This identity is called the {supersymmetric Ward identity}\index{supersymmetric Ward identity} \cite{Cecotti:1984wn}.


\section{Matter couplings in supergravity\protect\footnote{This section is reprinted from  \cite{DallAgata:2021uvl} © 2021 Springer-Verlag GmbH Germany, part of Springer Nature. Reproduced with permissions. All rights reserved.}}\label{cha:matter_couplings_in_supergraviy} 

Once the construction of the pure supergravity theory has been completed, we can try and analyze what needs to be done to couple matter multiplets in a consistent way.
This is what we are going to discuss in this section. In our presentation, we will limit as much as possible the details of the derivations and focus instead on the new features of matter couplings in supergravity as compared to what is already required by rigid supersymmetry.
In fact, already relaxing the requirement of studying renormalizable interactions generalizes quite a lot the possible couplings of globally supersymmetric theories, without the need of resorting to (super)gravity.
We would therefore like in the following to pinpoint the signatures that are unique to the supersymmetrization of the gravitational interaction.

The discussion on the matter couplings clearly depends heavily on what kind of matter we allow in these couplings and if we allow more than 2-derivative terms.
More general Lagrangians could be obtained by introducing additional matter multiplets, such as tensor multiplets.\index{tensor multiplets}
The tensor fields in such tensor multiplets, however, can in general be dualized to either massless scalar or massive vector fields so that the theory will be eventually of the standard form we will write down. 
On the other hand, these dualities are often non-perturbative or may require complicated field redefinitions. 
It may therefore be interesting to study these models directly with tensor fields, also because such tensor fields naturally arise from string theory compactifications, but we will not discuss this here.

Other generalizations may include higher degree form fields, higher spin fields and/or higher derivative terms.
While for each of these generalizations one can work out the construction that successfully leads to supersymmetric theories, we will stick to the simplest minimal theory coupled to vector and chiral multiplets, which are enough to provide rich theories with matter and gauge couplings.

\bigskip

In the following, we denote the complex scalars of $n_{C}$ chiral multiplets by $\phi^{m}$ ($m,n,\ldots=1,\ldots,n_{C}$) and their complex conjugates by $\phi^{\overline{m}}$ and use $\chi_{L}^{m}$ and $\chi_{R}^{\overline{m}}$ for the chiral projections of their fermionic superpartners. The component fields of $n_V$ vector multiplets are the gaugini, $\lambda^I$, and the vector fields, $A_{\mu}^{I}$  ($I,J,\ldots=1,\ldots,n_V$). 

General $\mathcal{N}=1$ globally supersymmetric theories of chiral and vector multiplets are completely specified by the following data 
\begin{itemize}
	\setlength\itemsep{2mm}
	\item The numbers, $n_{C}$ and $n_{V}$, of the chiral and vector multiplets;
	\item The K\"{a}hler potential $K(\phi^{m},\phi^{\overline{m}})$ that determines the geometry of the scalar manifold, $\mathcal{M}_{\textrm{scalar}}$; \index{K\"ahler potential}
	\item The holomorphic superpotential $W(\phi^m)$ that encodes the self-interactions of the chiral multiplets; \index{superpotential}
	\item The holomorphic gauge kinetic function $f_{IJ}(\phi^{m})$ related to the kinetic terms of the vector multiplets;\index{gauge kinetic function}
	\item The action of the gauge group on $\mathcal{M}_{\textrm{scalar}}$, as specified by the holomorphic Killing vectors,\index{Killing vectors} $\xi_{I}^{m}(\phi^{n})$, and the corresponding Killing prepotentials, $\mathcal{P}_{I}(\phi^{m},\phi^{\overline{m}})$;\index{Killing prepotentials}
	\item The real Fayet--Iliopoulos terms, $\eta_{I}$, which might be non-zero for \emph{Abelian} gauge group factors. \index{Fayet--Iliopoulos terms}
\end{itemize}
When we couple such a theory to supergravity, making it locally supersymmetric, there will be additional couplings of the matter multiplets to the supergravity multiplet, but also new and modified couplings among the fields of the matter multiplets themselves \cite{Cremmer:1982wb,Cremmer:1982en,Bagger:1982ab}. 
All these additional or modified couplings are still completely specified by the above-mentioned data that already specified a theory in global supersymmetry.
As we will now explain, they will appear in the Lagrangian with inverse powers of $M_{P}$.

\subsection{Coupling chiral multiplets to supergravity} 
\label{sub:coupling_chiral_multiplets_to_supergravity}

Let us start with the modifications that are necessary in order to make a globally supersymmetric Wess--Zumino model also invariant under local supersymmetry. \index{Wess--Zumino}
The globally supersymmetric Lagrangian in Minkowski space is 
\begin{eqnarray}
	\mathcal{L}_{WZ}&=&-g_{m\overline{n}}\left[(\partial_{\mu}\phi^{m})(\partial^{\mu}\phi^{\overline{n}}) + {\overline{\chi}}_{L}^{m} \slashed{\mathcal{D}} \chi_{R}^{\overline{n}} + {\overline{\chi}}_{R}^{\overline{n}} \slashed{\mathcal{D}} \chi_{L}^{m}\right]\nonumber\\[2mm]
	& & -(\mathcal{D}_{m}\partial_{n} W) \overline{\chi}{}_{L}^{m}\chi_{L}^{n}- (\mathcal{D}_{\overline{m}}\partial_{\overline{n}}W^{\ast})\overline{\chi}{}_{R}^{\overline{m}}\chi_{R}^{\overline{n}}\label{LWZ2} \\[2mm]
	& & - g^{m\overline{n}}(\partial_{m}W)(\partial_{\overline{n}} W^{\ast}) + \mathcal{O}(\chi^{4}) ,\nonumber 
\end{eqnarray}
where $g_{m\overline{n}} = \partial_{m}\partial_{\overline{n}}K$ is the K\"{a}hler metric. The derivative $\mathcal{D}$ is covariant with respect to arbitrary holomorphic scalar field reparameterizations and hence contains the Christoffel symbols, $\Gamma_{mn}^{p}$, on the scalar manifold, e.g.,
\begin{equation}
	\mathcal{D}_{\mu}\chi^{m}_{L}\equiv \partial_{\mu}\chi^{m}_{L} +(\partial_{\mu}\phi^{n})\Gamma_{nl}^{m}\chi^{l}_{L}.\label{chiDglobal} 
\end{equation}
If we now allow for a spacetime dependent supersymmetry parameter, $\epsilon=\epsilon(x)$, the derivative in the kinetic terms of the fermions $\chi^{m}$ will produce new terms when it acts on $\epsilon(x)$ coming from the supersymmetry transformations of the chiral fermions 
\begin{eqnarray}
	\delta \chi_{L}^{m}&=& \frac12 \slashed{\partial}\phi^{m}\epsilon_{R} -\frac12 g^{m\overline{n}}(\partial_{\overline{n}}W^{\ast})\epsilon_{L} + \mathcal{O}(\chi\chi\epsilon),\label{trafochi5}\\
	\delta \chi_{R}^{\overline{m}}&=& \frac12 \slashed{\partial}\phi^{\overline{m}}\epsilon_{L} -\frac12 g^{\overline{m}n}(\partial_{n}W)\epsilon_{R} +\mathcal{O}(\chi\chi\epsilon). 
	\label{trafochi6} 
\end{eqnarray}
The result is an uncancelled variation of the form 
\begin{equation}
	\delta \mathcal{L}_{\textrm{WZ}} = \overline J_{R}^{\mu}\, \partial_{\mu}\epsilon_{R}+\overline J_{L}^{\mu} \, \partial_{\mu}\epsilon_{L} ,
\end{equation}
where the supercurrents are
\begin{equation}
	\overline J_{L}^{\mu}=-g_{m\overline{n}}\overline{\chi}_{L}^{m}  \gamma^{\mu} \slashed{\partial}  \phi^{\overline{n}} + \overline{\chi}_{R}^{\overline{n}}\gamma^{\mu}\partial_{\overline{n}}W^\ast,\qquad J_{R}^{\mu}=(J_{L}^{\mu})^{c}.\label{Jdefinition} 
\end{equation}
As we have already shown in section \ref{sec:promoting_supersymmetry_to_a_local_symmetry} for the special case of a free Wess--Zumino model, the cancellation of these terms is achieved by adding the Noether couplings to the gravitino 
\begin{equation}
	\mathcal{L}_{\textrm{Noether}}=-\frac{1}{M_{P}}\left[ \overline J_{R}^{\mu}\psi_{\mu R} + \overline J_{L}^{\mu}\psi_{\mu L}\right].  \label{ChiralNoetherCoupling}
\end{equation}
Using $\delta \psi_{\mu}=M_{P}\, \partial_{\mu}\epsilon$, one then finds that everything cancels modulo terms that come from the variation of the supercurrents themselves: 
\begin{equation}
	\delta(\mathcal{L}_{WZ}+\mathcal{L}_{\textrm{Noether}})=-\frac{1}{M_{P}}\left[ (\delta \overline J_{R}^{\mu})\psi_{\mu R}+\textrm{ h.c. 
	}\right] 
\end{equation}
These terms are of the form 
\begin{equation}
	\delta(\mathcal{L}_{WZ}+\mathcal{L}_{\textrm{Noether}})=-\delta g^{\mu\nu}T_{\mu\nu} + Z_{1} + Z_{2}.\label{Z12def} 
\end{equation}
Just as discussed in chapter \ref{sec:promoting_supersymmetry_to_a_local_symmetry}, $T_{\mu\nu}$ is the energy momentum tensor of $\mathcal{L}_{WZ}$, and the new field $g_{\mu\nu}$ is identified with the spacetime metric, signalling the necessity for a coupling to gravity. The minimal coupling to a dynamical metric
is achieved by  covariantizing everything with respect to general spacetime coordinate and local Lorentz-transformations and by adding the pure supergravity Lagrangian. The metric variation of this covariantized Lagrangian then precisely cancels the first term in (\ref{Z12def}), and the theory would  be supersymmetric if there weren't also the two additional terms $Z_{1}$ and $Z_{2}$ in eq.~(\ref{Z12def}) that we have neglected so far. 
As we will now show, these two terms are actually quite important, as they lead to  additional $M_{P}^{-2}$-suppressed interactions between the fields of the chiral multiplets themselves that have
some far-reaching consequences.
  
In order to make this more precise, let us first state what $Z_{1}$ and $Z_{2}$ are:
\begin{eqnarray}
	Z_{1}&=&-\frac{e}{2M_{P}}\, g_{m\overline{n}}\,\overline{\psi}_{\mu}\gamma^{\mu\nu\rho}\gamma_{5}\epsilon\,(\partial_{\nu}\phi^{m})(\partial_{\rho}\phi^{\overline{n}}), \\[2mm]
	Z_{2}&=& \frac{e}{M_{P}}\,\left[\overline{\psi}_{\mu L}\gamma^{\mu\nu}\epsilon_{L}\,(\partial_{\nu} W^\ast) + \overline{\psi}_{\mu R}\gamma^{\mu\nu}\epsilon_{R}\,(\partial_{\nu} W)\right].
\end{eqnarray}
The first term $Z_{1}$ comes from the variations of the form $\delta \chi^{m}_{L}\sim \frac12 \slashed{\partial}\phi^{m}\epsilon_{R}$ in $J_{L}^{\mu}$ and its conjugate, which give rise to terms with three antisymmetrized gamma matrices as well as terms with one gamma matrix. 
The latter are part of the energy momentum tensor terms in (\ref{Z12def}) (because $\delta g_{\mu\nu}$ involves only one gamma matrix), whereas the terms with three antisymmetrized gamma matrices are precisely given by $Z_{1}$. 
The first term of $Z_{2}$ is due to the variations $\delta \chi^{m}_{L}\sim -\frac12 g^{m\overline{n}}(\partial_{\overline{n}}W^\ast)\epsilon_{L}$ in the first term in (\ref{Jdefinition}), and due to the variation $\delta \chi^{\overline{m}}_{R}\sim \frac12 \slashed{\partial}\phi^{\overline{m}}\epsilon_{L}$ in the second term in (\ref{Jdefinition}). 
The second term in $Z_2$ arises from the analogous variations of $J_R^{\mu}$.

We will now see that the cancellation of $Z_{1}$ and $Z_{2}$ requires the introduction of new terms with important consequences.


\subsection{The K\"{a}hler covariant derivative} 
\label{sub:the_k"_a_hler_covariant_derivative}

In order to cancel $Z_1$, we first rewrite it by using the relation between the metric of the scalar manifold and the K\"ahler potential:
\begin{equation}
	\begin{split}
	Z_{1} \sim &\overline {\psi}_{\mu}\gamma^{\mu\nu\rho} \gamma_{5} \epsilon\; \left(\partial_{\nu}\phi^{m}\right)\,\left(\partial_{\rho}\phi^{\overline{n}}\right)\, \partial_m  \partial_{\bar n} K \\[2mm]
	= &\overline{\psi}_{\mu}\gamma^{\mu\nu\rho}\gamma_{5} \epsilon \; \frac12 \,\left(\partial_\rho \phi^{\bar n}\partial_\nu  \partial_{\bar n} K -\partial_\rho \phi^m \partial_\nu \partial_m K\right).
	\end{split}
\end{equation}
Using the last expression, we then integrate by parts the spacetime derivative that acts on the K\"ahler potential. 
This produces in particular terms where the derivative acts on $\epsilon$ and terms where it acts on $\overline{\psi}_{\mu}$. 
The former term is
\begin{equation}
 \frac{e}{2M_P} \, \overline{\psi}_{\mu}\gamma^{\mu\nu\rho}\gamma_{5} (D_{\nu}\epsilon) \, \frac12 \left(\partial_\rho \phi^{\bar n}  \partial_{\bar n} K -\partial_\rho \phi^m \partial_m K\right).\label{tocancelthis}
 \end{equation}
We now repeat our old trick and simply add the negative of this term (times a factor 1/2) to the Lagrangian, but with 
 $D_\nu\epsilon$ replaced by $\psi_{\nu}$, 
\begin{equation}
	\mathcal{L}_{\textrm{K\"{a}hler cov}}= -\frac{e}{2}\overline{\psi}_{\mu}\gamma^{\mu\nu\rho}\left(\frac{i}{2M_{P}^{2}}Q_{\nu}(\phi)\gamma_{5}\right)\psi_{\rho},
\end{equation}
where $Q_\nu$ is a composite vector field,
\begin{equation}
	Q_{\nu}(\phi)\equiv \frac{i}{2}\left[(\partial_{\overline{n}}K)\partial_{\nu}\phi^{\overline{n}}-(\partial_{m}K)\partial_{\nu}\phi^{m}\right]. \label{Qnudef}
\end{equation}
Varying the two gravitini in this expression would then precisely cancel (\ref{tocancelthis}).

The cancellation of the remaining term in $Z_1$, where the derivative acts on $\overline{\psi}_{\mu}$, will be discussed later (see footnote \ref{dpsicancellation} in this section). 

The new interaction term $\mathcal{L}_{\textrm{K\"{a}hler cov}}$, however, now poses another problem: as one easily verifies, it is not invariant under K\"{a}hler transformations $K\rightarrow K+h+h^{\ast}$. 
$\mathcal{L}_{\textrm{K\"{a}hler cov}}$ would thus seem to single out a particular K\"{a}hler potential, even though a specific K\"{a}hler potential is not an intrinsic geometrical object on a K\"{a}hler manifold. 
In general, the K\"{a}hler potential is in fact only locally defined and requires K\"{a}hler transformations on the overlaps of local coordinate patches. 
So if the Lagrangian was not K\"{a}hler invariant, the physics would in general also be different for different coordinate patches of the scalar manifold.

To understand the resolution of this problem, we observe that the term $\mathcal{L}_{\textrm{K\"{a}hler cov}}$ can be absorbed into the Rarita--Schwinger action by modifying the covariant derivative with a new term,
\begin{equation}
	\mathcal{L}_{\textrm{RS}}+\mathcal{L}_{\textrm{K\"{a}hler cov}}= -\frac{e}{2}\overline{\psi}_{\mu}\gamma^{\mu\nu\rho}\mathcal{D}_{\nu}(\omega,Q)\psi_{\rho},\label{comb} 
\end{equation}
where \index{K\"ahler covariant derivative}
\begin{equation}
	\mathcal{D}_{[\nu}(\omega,Q)\psi_{\rho]}\equiv D_{[\nu}(\omega)\psi_{\rho]} +\frac{i}{2M_{P}^{2}}Q_{[\nu}\gamma_{5}\psi_{\rho]},\label{Kder} 
\end{equation}
with $D_{\nu}$ being the Lorentz covariant derivative. 
To understand the significance of this modification, one notes that $Q_\mu$ transforms under K\"{a}hler transformations like a U(1) connection:\index{composite K\"ahler connection}
\begin{equation}
	Q_\mu \rightarrow Q_\mu +  \partial_\mu \textrm{Im}(h).
	\label{QKahl}
\end{equation}
More precisely, $Q_{\mu}$ is a composite U(1) connection, i.e., it is not an elementary vector field, but rather a function of the scalar fields and their derivatives.

We now see that we can render the Lagrangian invariant if we require that  K\"{a}hler transformations, $K\rightarrow K+h+h^{\ast}$, be accompanied by chiral rotations of the gravitino:
\begin{equation}
	\psi_{\mu}\rightarrow \exp\left[-\frac{i}{2M_{P}^{2}}\textrm{Im}(h(\phi))\gamma_{5}\right]\psi_{\mu}. \label{psitrafo} 
\end{equation}
Indeed, the derivative (\ref{Kder}) then transforms covariantly, 
\begin{equation}
	\mathcal{D}_{[\mu} \psi_{\rho]}\rightarrow \exp\left[-\frac{i}{2M_{P}^{2}}\textrm{Im}(h(\phi))\gamma_{5}\right] \mathcal{D}_{[\mu}\psi_{\rho]},\label{Kepsilon} 
\end{equation}
and the combination (\ref{comb}) is K\"{a}hler-(and obviously also locally Lorentz-) invariant.
These geometric arguments thus suggest that, in supergravity, K\"ahler transformations on the scalar manifold also act on the gravitino as a chiral U(1) symmetry, with $Q_\mu$ being the corresponding (composite) U(1) connection.
If this is to make sense, this nontrivial action of K\"{a}hler transformations on the gravitini should also be compatible with supersymmetry. 
As we will now show, this requirement will lead to further interesting differences with respect to global supersymmetry and provides further consistency checks.

First we note that if the gravitino transforms under K\"{a}hler transformations, the consistency with the supersymmetry transformation law $\delta \psi_{\mu}\sim M_{P}D_{\mu}\epsilon$ also requires that $\epsilon$ transforms under K\"{a}hler transformations, 
\begin{equation}
	\epsilon \rightarrow \exp\left[-\frac{i}{2M_{P}^{2}}\textrm{Im}(h(\phi))\gamma_{5}\right]\epsilon \label{eT} 
\end{equation}
and that its derivative (as it appears in $\delta\psi_{\mu}$) should also be covariantized, \footnote{\label{dpsicancellation} This is indeed confirmed by computing the gravitino variations of $\mathcal{L}_{RS}$ with the  new K\"{a}hler covariant transformation law, $\delta \psi_{\mu}\sim M_{P}\mathcal{D}_{\mu}\epsilon$, which leads to a new term that precisely cancels the remaining uncancelled part of 
 $Z_1$ (i.e.~the part of $Z_{1}$ with a derivative acting on $\overline{\psi}_{\mu}$).}
\begin{equation}
	\mathcal{D}_{\mu}(\omega,Q)\epsilon\equiv D_{\mu}(\omega)\epsilon +\frac{i}{2M_{P}^{2}}Q_{\mu}\gamma_{5}\epsilon. 
	\label{epsilonder} 
\end{equation}
This in turn implies, because of $\delta\chi_{L}^{m}=\frac12 \slashed{\partial}\phi^{m}\epsilon_{R}+\ldots$, that also the chiral fermions transform under K\"{a}hler transformations, 
\begin{equation}
	\chi^{m}\rightarrow \exp\left[+\frac{i}{2M_{P}^{2}}\textrm{Im}(h(\phi))\gamma_{5}\right]\chi^{m},\label{chiK} 
\end{equation}
and that their derivatives have to be Lorentz-, $\mathcal{M}_{\textrm{scalar}}$-reparameterization- and K\"{a}hler-covariant, e.g., \footnote{For the sake of simplicity, we do not introduce a new symbol for the K\"{a}hler covariantized derivative and still call it $\mathcal{D}_{\mu}$.} 
\begin{equation}
	\mathcal{D}_{\mu}\chi^{m}_{L}\equiv D_{\mu}\chi^{m}_{L} +(\partial_{\mu}\phi^{n})\Gamma_{nl}^{m}\chi^{l}_{L} -\frac{i}{2M_{P}^{2}}Q_{\mu}\chi^{m}_{L}.\label{chiD} 
\end{equation}
Note that there is a different sign in (\ref{chiK}) (and hence also in (\ref{chiD})) compared to the corresponding terms of the gravitino or the supersymmetry transformation parameter (cf.~(\ref{psitrafo}) and (\ref{eT}) as well as (\ref{Kder}) and (\ref{epsilonder})). 
This sign difference arises because one has to move the $\gamma_{5}$ matrix in (\ref{eT}) through one gamma matrix in the supersymmetry transformation $\delta\chi_{L}^{m}=\frac12 \slashed{\partial}\phi^{m}\epsilon_{R}+\ldots$. 

Although we will discuss gauge multiplets later, we already mention here that $\delta \lambda^{I} \sim \frac14 \gamma^{\mu\nu}\mathcal{F}_{\mu\nu}^{I}\epsilon+\ldots$ implies that also the gaugini transform nontrivially under K\"{a}hler transformations (with the same sign as $\psi_{\mu}$ and $\epsilon$) 
\begin{equation}
	\delta \lambda^I \rightarrow \exp\left[-\frac{i}{2M_{P}^{2}}\textrm{Im}(h(\phi))\gamma_{5}\right] \lambda^I,
	\label{chilambda}
\end{equation}
and that likewise all their derivatives have to be properly covariantized with respect to K\"{a}hler transformations (again with the same sign as for $\psi_{\mu}$ and $\epsilon$).

To conclude, all fermion fields and not just the gravitino are charged with respect to a composite chiral U(1) symmetry that is related to K\"{a}hler transformations and that is not present in the global case.
It should be emphasized that in the limit of global supersymmetry, $M_P \to \infty$, these chiral rotations become trivial, as is signalled by the inverse powers of $M_P$.
This is consistent with the rigid supersymmetry Lagrangian (\ref{LWZ2}), where this chiral composite U(1) is not encountered.

Interestingly, the above non-trivial transformations of the fermions under K\"{a}hler transformations also imply that the superpotential and its derivatives have to transform as we will show in section \ref{sub:additional_bare_superpotential_terms} below.
The result is that 
\begin{equation}
	W\rightarrow \exp\left[-\frac{1}{M_{P}^{2}}h(\phi^{m})\right]W(\phi^{m})\label{WT} 
\end{equation}
and its derivatives have to be K\"{a}hler covariantized as follows
\begin{equation}
	\partial_{n}W\rightarrow e^{\frac{K}{2M_{P}^{2}}}\mathcal{D}_{n}W\equiv e^{\frac{K}{2M_{P}^{2}}}\left[\partial_{n} + \frac{(\partial_{n} K)}{M_{P}^{2}}\right]W \label{Wd}.
\end{equation}

To summarize: The cancellation of $Z_{1}$ by adding $\mathcal{L}_{\textrm{K\"{a}hler cov}}$ gives rise to the interpretation that the fermions and the superpotential should transform non-trivially under K\"{a}hler transformations. 
In order to ensure this, all derivatives of the fermions and the superpotential have to be K\"{a}hler covariantized, and the superpotential terms have to be dressed with an exponential of the K\"{a}hler potential. 
One can show that all these modifications are indeed also necessary for the cancellation of various other variations we have not discussed here in detail. 
In general, we define the K\"ahler covariant derivatives in field space as\index{K\"ahler covariant derivative}
\begin{equation}
	\label{covDbundle}
	\begin{split}
		D_m \Phi &= \left(\partial_m + \frac{p}{M_P^2} \partial_m K\right) \Phi, \\[2mm]	
		\overline D_{\overline m} \Phi &= \left(\overline{\partial}_{\overline m} - \frac{p}{M_P^2} \overline{\partial}_{\overline m} K \right) \Phi, 
	\end{split}
\end{equation}
where $p$ is the K\"ahler ``charge'' of the field $\Phi$.


\subsection{Additional bare superpotential terms} 
\label{sub:additional_bare_superpotential_terms}

In global supersymmetry, all superpotential terms always appear with at least one derivative with respect to the scalar fields. 
As we saw in the previous subsection, the coupling to supergravity (in particular the cancellation of the term $Z_{1}$) requires a K\"{a}hler covariantization of these derivatives of $W$, which then introduces ``bare'' $W$-terms inside these K\"{a}hler covariant derivatives, i.e., $W$-terms that are not differentiated with respect to any scalar field. 
In this subsection, we show that there are additional ``bare'' superpotential terms in the Lagrangian and the supersymmetry transformation laws. 
Their necessity follows from the cancellation of the term $Z_{2}$ to which we now turn.\index{superpotential}

In order to cancel the $Z_{2}$-term,
\begin{equation}
	Z_{2}\equiv \frac{e}{M_{P}}\left[\overline{\psi}_{\mu L}\gamma^{\mu\nu}\epsilon_{L}(\partial_{\nu} W^{\ast}) + \overline{\psi}_{\mu R}\gamma^{\mu\nu}\epsilon_{R}(\partial_{\nu} W)\right],
\end{equation}
we proceed as we did for $Z_1$ and first perform an integration by parts. 
This will then give again terms with a derivative acting on the supersymmetry parameter $\epsilon$ and terms where the derivative acts on on the gravitini $\overline{\psi}_{\mu}$. 
To cancel the former, we then again add to the Lagrangian a term where the derivatives of $\epsilon$ are replaced by gravitini, or, more precisely,
\begin{equation}
	\frac{e}{2M_{P}^{2}} \,\left[ W^{\ast} \,\overline{\psi}_{\mu L}\gamma^{\mu\nu}\psi_{\nu L} +  \, {W} \,\overline{\psi}_{\mu R}\gamma^{\mu\nu}\psi_{\nu R}\right]. \label{WW} 
\end{equation}
This term is an obvious mass-like term for the gravitino and therefore, following the rules we have learned in the case of pure supergravity in the presence of a cosmological constant, we have to further modify the variation of the gravitino field by adding a new term of the form
\begin{equation}
	\delta_{\textrm{new}}\psi_{\mu L}\sim \frac{1}{2M_{P}}\, W\,\gamma_{\mu}\epsilon_{R}. \label{WWW} 
\end{equation}
This new variation applied to the Rarita--Schwinger action also gives the term required to cancel the second piece coming from the partial integration of $Z_2$, namely the term with the derivative acting on the gravitino.

Before proceeding further, let us come back to the K\"ahler covariantization of the superpotential terms and prove (\ref{WT}).
Subjecting (\ref{WWW}) to K\"{a}hler transformations tells us that the left hand side transforms as 
\begin{equation}
	\exp\left[-\frac{1}{4M_{P}^{2}}(h(\phi)-h^*(\phi))\right],
\end{equation}
while the epsilon parameter on the right hand side transforms with the opposite sign due to the opposite chirality:
\begin{equation}
	\exp\left[+\frac{1}{4M_{P}^{2}}(h(\phi)-h^*(\phi))\right].
\end{equation}
At this point it is obvious that in order for the K\"{a}hler transformation to be compatible with supersymmetry, we need to transform also the superpotential, as we already mentioned earlier.
The superpotential, on the other hand, is a \emph{holomorphic} function by construction and hence can transform only with a holomorphic factor,
\begin{equation}
	W \to \exp\left[-\frac{\alpha}{M_{P}^{2}}(h(\phi))\right]W,
\end{equation}
where $\alpha$ is a real constant.
In order to get the same rotation on the left and on the right hand side of (\ref{WWW}) we still need something that transforms under K\"ahler transformations with the exponential of $h + h^*$, like the exponential of the K\"ahler potential itself, ${\rm e}^{\beta K/M_P^2}$.
The right coefficients follow then by equating the two sides:
\begin{equation}
-\frac{1}{4M_{P}^{2}}(h(\phi)-h^*(\phi))= +\frac{1}{4M_{P}^{2}}(h(\phi)-h^*(\phi)) - \frac{\alpha}{M_{P}^{2}}(h(\phi)) + \frac{\beta}{M_P^2} (h(\phi)+h^*(\phi)).
\end{equation}
This fixes $\alpha= 1$, $\beta=1/2$ and tells us that we have to replace the superpotential with the combination 
\begin{equation}\label{eKW}
	{\rm e}^{K/(2 M_P^2)} W
\end{equation}
and that indeed $W$ transforms under K\"{a}hler transformations as in (\ref{WT}).

Coming back to the check of supersymmetry invariance, we now see that the new transformation law for the gravitino (\ref{WWW}) applied to the new bilinear term (\ref{WW}), gives a new variation of the form $|W|^{2}\overline{\psi}\gamma\epsilon$. 
Not too surprisingly, this can then finally be cancelled by adding a new contribution $\sim -e|W|^{2}$ to the scalar potential and varying the vierbein determinant $e$. 
This is the generalization of the procedure derived in section \ref{sec:adding_a_cosmological_constant} for the case of a constant superpotential, i.e., for pure supergravity with a cosmological constant.

Although it may be hard to believe, it turns out that, after proper K\"{a}hler covariantizations, the above modifications are sufficient to ensure also the cancellations of all the other variations we have not considered explicitly here.

The end result is the Lagrangian  
\begin{eqnarray}
	e^{-1}\mathcal{L}&=& \frac{M_{P}^{2}}{2}R(e,\omega(e)) -\frac{1}{2} {\overline{\psi}}_{\mu}\gamma^{\mu\nu\rho}\mathcal{D}_{\nu}(\omega(e),Q)\psi_{\rho}\nonumber\\[2mm]
	&& -g_{m\overline{n}} \left[ (\partial_{\mu}\phi^{m})(\partial^{\mu}\phi^{\overline{n}}) + {\overline{\chi}}_{L}^{m} \slashed{\mathcal{D}} \chi_{R}^{\overline{n}} + {\overline{\chi}}_{R}^{\overline{n}} \slashed{\mathcal{D}} \chi_{L}^{m} \right]\nonumber\\[2mm]
	&&- \Big\{e^{K/2M_{P}^{2}}(\mathcal{D}_{m}\mathcal{D}_{n} W) {\overline{\chi}}^{m}_{L}\chi^{n}_{L} + \textrm{h.c.} \Big\} \label{completec} 	 \\
	&& + \frac{1}{M_{P}}\Big\{g_{m\overline{n}}{\overline{\psi}}_{\mu L}\gamma^{\nu}\gamma^{\mu}\chi_{L}^{m} (\partial_{\nu}\phi^{\overline{n}}) + {\overline{\psi}}_{\mu R}\gamma^{\mu}\chi_{L}^{m} e^{K/2M_{P}^{2}}\mathcal{D}_{m}W +\textrm{h.c.} \Big\} \nonumber\\[2mm]
	&& + \frac{1}{2M_{P}^{2}} \Big\{ e^{K/2M_{P}^{2}} W {\overline{\psi}}_{\mu R}\gamma^{\mu\nu} \psi_{\nu R} + h.c \Big\} -V(\phi^{m},\phi^{\overline{n}}), \nonumber
\end{eqnarray}
with the scalar potential given by the sum of two contributions
\begin{equation}
	V= e^{K/M_{P}^{2}} \left[  g^{m\overline{n}}(\mathcal{D}_{m}W) (\mathcal{D}_{\overline{n}} W^\ast) -\frac{3|W|^{2}}{M_{P}^{2}}\right], \label{potc} 
\end{equation}
where the first term is the K\"ahler covariantization of the F-terms from global supersymmetry and the second is a genuine contribution from gravitational couplings, in the sense that it is a variation of the vierbein determinant that leads to a cancellation of the $|W|^2$ terms mentioned after (\ref{eKW}).
In the Lagrangian (\ref{completec}), the first line is the K\"{a}hler covariantization of the pure supergravity action. 
The second and third line correspond to the K\"{a}hler and spacetime covariant Wess--Zumino action (without the potential). 
Note that now
\begin{equation}\label{expansioncovariants}
		{\cal D}_m {\cal D}_n W = \left(\partial_m + \frac{\partial_m K}{M_P^2}\right)\left[\left(\partial_n + \frac{\partial_n K}{M_P^2}\right) W\right] - \Gamma_{mn}{}^p \left(\partial_p + \frac{\partial_p K}{M_P^2}\right)W.
\end{equation}
The fourth line is the K\"{a}hler and spacetime covariant Noether coupling of the supercurrents to the gravitino, $\mathcal{L}_{\textrm{Noether}}=-\frac{1}{M_{P}}\left[\overline{J}_{R}^{\mu}\psi_{\mu R}+ \overline{J}_{L}^{\mu}\psi_{\mu L}\right]$ (with the fermions moved into a different order). 
The fifth line, finally, contains the $W$-dependent extra terms as well as the (K\"{a}hler covariantized) scalar potential of the Wess--Zumino model.

The supersymmetry transformation rules, up to 3-fermion terms, are 
\begin{equation}\label{cT}
	\begin{array}{rcl}
	\delta e_{\mu}^{a} &=& \displaystyle \frac{1}{2M_{P}} \overline{\epsilon} \gamma^{a}\psi_{\mu}, \\[2mm]
	\delta \psi_{\mu L} & = &\displaystyle M_{P}\mathcal{D}_{\mu}(\omega(e), Q_{\nu}) \epsilon_{L} +\frac{1}{2M_{P}} e^{K/2M_{P}^{2}}W \gamma_{\mu}\epsilon_{R}, \\
	\delta \phi^{m}&=& {\overline{\epsilon}}_{L}\chi^{m}_{L},\\[2mm]
	\delta \chi^{m}_{L} & = &\displaystyle \frac{1}{2} \slashed{\partial} \phi^{m} \epsilon_{R}-\frac{1}{2} g^{m\overline{n}}e^{K/2M_{P}^{2}}(\mathcal{D}_{\overline{n}} W^{\ast}) \epsilon_{L}. 
\end{array}
\end{equation}
Obviously, in the $M_{P} \rightarrow \infty$ limit, these equations reduce to the globally supersymmetric theory. 
One also notices that truncating out the chiral multiplets and keeping a constant superpotential ${\rm e}^{\frac{K}{2M_P^2}}W= -g \, M_{P}^{3}$ gives back the pure supergravity Lagrangian with cosmological constant, eq.~(\ref{SugraLambda}).

Note further that in supergravity the K\"{a}hler potential and the superpotential are no longer independent, as one can shift terms back and forth via K\"{a}hler transformations. 
In fact, as long as $W$ is not equal to zero, one can even make the superpotential equal to $M_{P}^{3}$ by performing a K\"{a}hler transformation with $h(\phi)=M_{P}^{2} \,\log (W/M_{P}^{3})$ (cf.~eq.~(\ref{WT})). 
More generally, instead of using the two functions $K$ and $W$, one can express the entire Lagrangian in terms of the function \index{K\"ahler potential}
\begin{equation}\label{Gpotential}
	\mathcal{G}=K+M_{P}^{2}\; \log \frac{|W|^2}{M_{P}^{6}}, 
\end{equation}
which is manifestly K\"{a}hler invariant. 
For instance, the part of the scalar potential coming from the superpotential becomes 
\begin{equation}\label{Gpotential2}
		V = e^{{\cal G}/M_P^2}\left(M_P^2\,g^{m \bar{n}} {\cal G}_m {\cal G}_{\bar n} - 3 M_P^4\right).
\end{equation}
Note, however, that by doing so one cannot recover the $W=0$ case, which has to be discussed separately.
Hence the usefulness of leaving explicit both $K$ and $W$ in our approach.


\subsection{Inclusion of vector multiplets} 
\label{sub:inclusion_of_vector_multiplets}

The inclusion of vector multiplets requires the following changes: \index{vector multiplet}
\begin{enumerate}
	\setlength\itemsep{2mm}	
	\item All terms that were already present in global supersymmetry are also present in supergravity, but they all have to be made spacetime and K\"{a}hler covariant. 
	\item A new Noether coupling of the vector multiplet supercurrent 
\begin{equation}
	\overline{J}^{\mu}_{VM}\equiv e\overline{\lambda}^{J}\left[-\frac{1}{4}(\textrm{Re} f_{IJ})\mathcal{F}_{\nu\rho}^{I}\gamma^{\mu}\gamma^{\nu\rho}-\frac{i}{2}\mathcal{P}_{J}\gamma^{\mu}\gamma_{5}\right] 
\end{equation}
to the gravitino has to be introduced: 
\begin{equation}
	\mathcal{L}_{\textrm{Noether}}^{\prime}=-\frac{1}{M_{P}} \overline{J}^{\mu}_{VM}\psi_{\mu} 
\end{equation}
in order to cancel terms of the form 
\begin{equation}
	\delta \mathcal{L}=\overline{J}_{VM}^{\mu}\partial_{\mu}\epsilon 
\end{equation}
that arise due to the derivative in $-\frac12 e\, (\textrm{Re} f_{IJ}) \overline{\lambda}^{I}\widehat{\slashed{\partial}}\lambda^{J}$ when it acts on the $\epsilon$ in $\delta \lambda$. 
In the above equations, $\mathcal{F}_{\mu\nu}^{I}$ denotes the usual gauge covariant field strengths, and a hat on a derivative denotes a gauge covariant derivative.

	\item The composite K\"{a}hler connection\index{composite K\"ahler connection} $Q_{\mu}$ receives an additional contribution proportional to $A_{\mu}^{I}\mathcal{P}_{I}$ for each of the gauged isometries: 
\begin{eqnarray}
	\hspace{-4mm} Q_{\mu}=Q_{\mu}(\phi^{m},\phi^{\overline{n}},A_{\mu}^{I})&=& \frac{i}{2} \left((\partial_{\overline{n}} K) \partial_{\mu}\phi^{\overline{n}} -(\partial_{m} K) \partial_{\mu} \phi^{m}\right) + A_{\mu}^{I}\mathcal{P}_{I}\\
	&=& \frac{i}{2} \left((\partial_{\overline{n}} K) \widehat{\partial}_{\mu}\phi^{\overline{n}} -(\partial_{m} K) \widehat{\partial}_{\mu} \phi^{m}\right) +A_{\mu}^I\textrm{Im}(r_{I}), \label{gbgb} 
\end{eqnarray}
where the last equality follows from the form of the prepotentials (we will see more on this in section \ref{sub:more_on_d_terms}).
This additional term is needed, e.g., in order to cancel a variation proportional to $\mathcal{F}_{\mu\nu}^{J}\mathcal{P}_{J} \overline{\epsilon}\gamma^{\mu\nu\rho}\gamma_{5}\psi_{\rho}$ that occurs in the variation $-\frac{1}{M_{P}}\delta(J_{VM}^{\mu})\psi_{\mu}$ and is not of the form $-\delta g^{\mu\nu}T_{\mu\nu}$. 

It should be noted that this additional contribution to $Q_{\mu}$ has another important consequence. 
Namely, if one shifts the Killing prepotential $\mathcal{P}_{I}$ of an Abelian factor by a Fayet--Iliopoulos constant, $\mathcal{P}_{I}\rightarrow \mathcal{P}_{I}+\eta_{I}$, one introduces new chiral gauge interactions for all fermions, including, e.g., the gravitino, 
\begin{equation}
	\mathcal{D}_{\mu}\psi_{\nu}\rightarrow \mathcal{D}_{\mu}\psi_{\nu}+ \frac{i}{2M_{P}^{2}}A_{\mu}^{I}\eta_{I}\gamma_{5}\psi_{\nu} , \label{anomalyr}
\end{equation}
which can easily lead to quantum anomalies \cite{Elvang:2006jk}. 
Thus, the introduction of Fayet--Iliopoulos constants in $\mathcal{N}=1$ supergravity requires some care. 
We will actually see later on that in supergravity the Fayet--Iliopoulos terms are related to the non-invariance of the superpotential under gauge transformations.
\end{enumerate}
Ignoring 4-fermion terms, the end result of all these modifications is the following general matter coupled Lagrangian \footnote{When the gauge kinetic function is not gauge invariant, so-called generalized Chern-Simons terms of the form $A^{I}\wedge A^{J}\wedge dA^{K}$ and $A^{I}\wedge A^{J}\wedge A^{K}\wedge A^{L}$ may be possible. 
Their form, however, is the same as in global supersymmetry \cite{DeRydt:2007vg}.} 
\begin{eqnarray}
	e^{-1}\mathcal{L}&=& \frac{M_{P}^{2}}{2}R(e,\omega(e)) -\frac{1}{2} {\overline{\psi}}_{\mu}\gamma^{\mu\nu\rho}\mathcal{D}_{\nu}(\omega(e),Q)\psi_{\rho} \nonumber \\
	&&-g_{m\overline{n}} \left[ ({\widehat{\partial}}_{\mu}\phi^{m})({\widehat{\partial}}^{\mu}\phi^{\overline{n}}) + {\overline{\chi}}_{L}^{m} \not\!\! {\widehat{\mathcal{D}}} \chi_{R}^{\overline{n}} + {\overline{\chi}}_{R}^{\overline{n}} \not\!\! {\widehat{\mathcal{D}}} \chi_{L}^{m} \right]\nonumber\\
	& & +(\textrm{Re} f_{IJ}) \left[ -\frac{1}{4} \mathcal{F}_{\mu\nu}^{I}\mathcal{F}^{\mu\nu \, J} - \frac{1}{2} {\overline{\lambda}}^{I} \not\!\! {\widehat{\mathcal{D}}} \lambda^{J} \right] \nonumber \\
	&& + \frac{1}{8} (\textrm{Im} f_{IJ}) \left[ \mathcal{F}_{\mu\nu}^{I}\mathcal{F}^{J}_{\rho\sigma} \epsilon^{\mu\nu\rho\sigma} -2i\, {\widehat{\mathcal{D}}}_{\mu} (e{\overline{\lambda}}^{I}\gamma_{5}\gamma^{\mu}\lambda^{J}) \right] \nonumber\\
	&& + \left\{ -\frac{1}{4}f_{IJ,m}{\mathcal{F}}_{\mu\nu}^{I}{\overline{\chi}}^{m}_{L}\gamma^{\mu\nu}\lambda_{L}^{J} +\frac{i}{2}D^{I}f_{IJ,m}{\overline{\chi}}^{m}_{L}\lambda^{J}  \right.\nonumber \\
	&& +\frac{1}{4} e^{K/2M_{P}^{2}} (\mathcal{D}_{m} W)g^{m\overline{n}}f^{\ast}_{IJ,\overline{n}}{\overline{\lambda}}^{I}_{R}\lambda^{J}_{R} \nonumber \\
	&& \left.-e^{K/2M_{P}^{2}}(\mathcal{D}_{m}\mathcal{D}_{n} W) {\overline{\chi}}^{m}_{L}\chi^{n}_{L} -2\xi_{I}^{\overline{n}}g_{m\overline{n}}{\overline{\lambda}}^{I}\chi^{m}_{L} + \textrm{h.c.} \right\} \nonumber \\
	&& + \frac{1}{4M_{P}}(\textrm{Re}f_{IJ}){\overline{\psi}}_{\mu}\gamma^{\nu\rho}\gamma^{\mu}\lambda^{J} \mathcal{F}_{\nu\rho}^{I} \nonumber\\
	&& + \left\{ \frac{1}{M_{P}}g_{m\overline{n}}{\overline{\psi}}_{\mu L}\gamma^{\nu}\gamma^{\mu}\chi_{L}^{m} ({\widehat{\partial}}_{\nu}\phi^{\overline{n}}) + h.c. 
	\right\} \nonumber \\
	&&+\frac{1}{M_{P}}\left\{ {\overline{\psi}}_{\mu R}\gamma^{\mu} \left[ \frac{i}{2}\lambda^{I}_{L}\mathcal{P}_{I} + \chi_{L}^{m} e^{K/2M_{P}^{2}}\mathcal{D}_{m}W \right] +h.c. 
	\right\} \nonumber\\
	&& + \frac{1}{2M_{P}^{2}} \left\{ e^{K/2M_{P}^{2}} W {\overline{\psi}}_{\mu R}\gamma^{\mu\nu} \psi_{\nu R} + h.c \right\} -V(\phi^{m},\phi^{\overline{n}}), \label{complete} 
\end{eqnarray}
with the scalar potential 
\begin{equation}
	V=e^{K/M_{P}^{2}} \left[ g^{m\overline{n}}(\mathcal{D}_{m}W) (\mathcal{D}_{\overline{n}} \overline{W}) -\frac{3|W|^{2}}{M_{P}^{2}} \right] + \frac{1}{2} (\textrm{Re} f_{IJ} ) D^{I} D^{J}. 
	\label{pot} 
\end{equation}
The supersymmetry transformation rules, up to 3-fermion terms, are 
\begin{eqnarray}
	\delta e_{\mu}^{a} &=& \frac{1}{2M_{P}} \overline{\epsilon} \gamma^{a}\psi_{\mu}, \nonumber\\
	\delta \psi_{\mu L} & = & M_{P}\mathcal{D}_{\mu}(\omega(e), Q) \epsilon_{L} +\frac{1}{2M_{P}} e^{K/2M_{P}^{2}}W \gamma_{\mu}\epsilon_{R}, \nonumber\\
	\delta \phi^{m}&=& {\overline{\epsilon}}_{L}\chi^{m}_{L},\nonumber\\
	\delta \chi^{m}_{L} & = & \frac{1}{2} \not\! {\widehat{\partial}} \phi^{m} \epsilon_{R}-\frac{1}{2} g^{m\overline{n}}e^{K/(2M_{P}^{2})}(\mathcal{D}_{\overline{n}} W^\ast) \epsilon_{L},\nonumber \\
	\delta A_{\mu}^{I} & = & - \frac{1}{2}{\overline{\epsilon}}\gamma_{\mu}\lambda^{I}, \nonumber\\
	\delta \lambda^{I} &=& \frac{1}{4}\gamma^{\mu\nu}{\mathcal{F}}^{I}_{\mu\nu}\epsilon + \frac{i}{2}\gamma_{5}D^{I} \epsilon. 
\end{eqnarray}
It is again easy to see that, in the global limit, $M_{P} \rightarrow \infty$, the above equations reduce to the globally supersymmetric theory.

For completeness, we display the full (i.e., local Lorentz-, scalar coordinate-, K\"{a}hler-, and gauge- covariant) derivative of $\lambda^{I}$ and $\chi^{m}$: 
\begin{eqnarray}
	{\widehat{\mathcal{D}}}_{\mu} \chi^{m}_{L} &=& D_{\mu}\chi^{m}_{L}+({\widehat{\partial}}_{\mu}\phi^{n}) \Gamma_{nl}^{m}\chi_{L}^{l} -A_{\mu}^{I}(\partial_{n}\xi_{I}^{m})\chi_{L}^{n} -\frac{i}{2M_{P}^{2}}Q_{\mu} \chi^{m}_{L},\nonumber\\
	{\widehat{\mathcal{D}}}_{\mu} \lambda^{I} &=& D_{\mu}\lambda^{I} +A_{\mu}^{J}f_{JK}^{I}\lambda^{K} +\frac{i}{2M_{P}^{2}}Q_{\mu}\gamma_{5} \lambda^{I}, 
\end{eqnarray}
where, as usual, $D_{\mu}$ denotes the Lorentz covariant derivative. 
The full covariant derivatives of $\psi_{\mu}$ and $\epsilon$ are just as for $\lambda^{I}$, except for the gauge covariantization term $A_{\mu}^{J}f_{JK}^{I}\lambda^{K}$, which is absent for these fermions (hence, we can omit the hat on their derivatives).

More details on the action and the 4-Fermi terms can be found in \cite{Binetruy:2004hh,DeRydt:2007vg}.


\subsection{More on D-terms} 
\label{sub:more_on_d_terms}
Although the D-terms and the D-term potential take the same form as in global supersymmetry,\index{D-term}
local supersymmetry does have some interesting implications also for the D-terms. 
To understand this, we recall that the general matter-coupled supergravity Lagrangian is invariant under K\"ahler transformations that act at the same time on the K\"ahler potential, the superpotential and the fermions.
As in global supersymmetry, a gauge transformation therefore does not necessarily have to leave the K\"{a}hler potential invariant, but may in general transform it with a K\"ahler transformation, 
\begin{equation}
	\delta_{\rm{gauge}} K     \equiv \xi_{I}^{m}\partial_{m}K+\xi_{I}^{\overline{m}}\partial_{\overline{m}}K= r_I + r_I^\ast. \label{Ktrafo2}
\end{equation}
However, in supergravity theories this requires a non-trivial action also on the superpotential (if $W\neq 0$)
\begin{equation}
	\delta_{\rm{gauge}} W \equiv \xi_{I}^{m}\partial_{m}W= - \frac{r_I}{M_P^2} W,
	\label{dgaugeW}
\end{equation}
so that the combination ${\cal G}$ in (\ref{Gpotential}) remains invariant:
\begin{equation}
	\delta_{\rm{gauge}} {\cal G} = \xi_{I}^{m}\partial_{m}\mathcal{G} + {\xi}_{I}^{\bar{m}}{\partial}_{\bar{m}}\mathcal{G} = 0
\end{equation}
For all the points in field space where $W\neq 0$ we can then also rewrite $r_{I}$ as 
\begin{equation}
r_{I}=-M_{P}^{2}\,\xi^{m}_{I}\frac{\partial_{m}W}{W},	
\end{equation}
so that the Killing prepotentials can be also expressed in terms of the gauge-invariant quantity ${\cal G}$:
\begin{equation}
\mathcal{P}_{I}=i\xi_{I}^{m}\partial_{m}K-ir_{I}=
i\xi_{I}^{m}\left[\partial_{m} K +\frac{M_P^2 \partial_{m} W}{W}\right]=i\xi_{I}^{m}\frac{M_{P}^{2}\mathcal{D}_{m}W}{W}= i\xi_{I}^{m}\partial_{m}\mathcal{G}.	\label{DFterms}
\end{equation}
The D-terms are thus
\begin{equation}
D^I=i\, (\textrm{Re }f)^{-1 IJ}\,\xi_{J}^{m}\,\partial_{m}\mathcal{G},\label{DFtermsrelation}
\end{equation}
and the total scalar potential with F-terms and D-terms can be written in a very compact and suggestive form
\begin{equation}
V=e^{\mathcal{G}/M_{P}^{2}}\left[h^{m\overline{n}}\, \mathcal{G}_{m}\mathcal{G}_{\overline{n}}-3\right] M_P^2,	
\end{equation}
where
\begin{equation}
h^{m\overline{n}}\equiv g^{m\overline{n}}+\frac{e^{-\mathcal{G}/M_{P}^{2}}}{2}\left(\textrm{Re } f\right)^{-1 IJ}\xi_{I}^{m}\xi_{J}^{\overline{n}},
\end{equation}
so that the new metric contains the K\"ahler metric giving the F-term potential (\ref{Gpotential2}) and the additional term coming from the D-term potential.\index{D-term SUSY breaking}

We can now comment on some of the differences between global and local supersymmetry, which may lead to relevant physical differences.

First of all, we see that in supergravity D-terms and F-terms are not independent of one another, rather the D-terms are (for $W\neq 0$) a particular combination of the F-terms (\ref{DFtermsrelation}). 

The next important difference concerns the Fayet--Iliopoulos constants.
Just as in global supersymmetry, the gauge transformation (\ref{Ktrafo2}) of the K\"{a}hler potential fixes $r_{I}$ only up to an additive imaginary constant, $i\eta_{I}$ and hence $\mathcal{P}_{I}$  up to an additive real constant $\eta_{I}$. 
However, consistency of the $\mathcal{P}_I$ moment maps with the gauge symmetry requires the equivariance condition
\begin{equation}
	\{\mathcal{P}_I, \mathcal{P}_J\} = \frac12\, (\delta_I \mathcal{P}_J - \delta_J \mathcal{P}_I) = f_{IJ}{}^K \mathcal{P}_K
\end{equation}
and this restricts the possible values of the $\eta_I$ constants except for U(1) factors.   
The difference to global supersymmetry now is that the superpotential $W$ also transforms under gauge transformations as in (\ref{dgaugeW}) so that a shift of $r_{I}$ by an additive constant $i\eta_{I}$ implies that $W$ transforms with an additional phase factor under the corresponding U(1) transformation. 
In other words, changing $\eta_I$ changes the U(1) charge of $W$. 
Note that such U(1) transformations due to FI constants may even occur when the K\"ahler potential is invariant under this U(1) factor, because, according to (\ref{Ktrafo2}), this only implies $r_{I}(\phi)=i\,\eta_{I}$.

Another important effect of a FI constant is that it leads to a chiral U(1) transformation of the fermions, as follows from their non-trivial transformations under 
K\"{a}hler transformations described in section  \ref{sub:the_k"_a_hler_covariant_derivative}.  As explained around (\ref{anomalyr}), this may then easily lead to anomalous 
gauge couplings and requires some care.



\begin{acknowledgement}
This review is based on the book \cite{DallAgata:2021uvl}, which grew out of the many interactions we had in our careers. We would only like here to recall those who played a pivotal role in shaping our understanding of supergravity: A. Ceresole, R. D'Auria, S. Ferrara, M. G\"unaydin, A. Van Proeyen and F. Zwirner. We also thank once more L. \'Alvarez Gaum\'e, who first suggested to turn our lecture notes into a book.
\end{acknowledgement}


\section{Appendix: Conventions, spinors and useful relations}
\label{sec:spinors}

\subsection{General Relativity and spacetime conventions}
We use $x^\mu$ ($\mu,\nu,\ldots=0,1,2,3$) to denote local 4D spacetime coordinates and $\partial_{\mu}$ as the corresponding coordinate basis vectors of the tangent spaces. 

The metric tensor has components $g_{\mu\nu}$ and signature $(-+++)$, and the Christoffel symbols, $\Gamma_{\mu\nu}^{\rho}=\Gamma_{\nu\mu}^{\rho}$, always refer to the usual torsion-free Levi-Civita connection,
\begin{equation}
	\Gamma_{\mu\nu}^\rho =\Gamma_{\mu\nu}^\rho (g)= \frac12 g^{\rho\sigma}\left(
	\partial_\mu g_{\nu\sigma} + 
	\partial_\nu g_{\mu\sigma} - 
	\partial_\sigma g_{\mu\nu}\right).
\label{LeviCivita} 
\end{equation}
The curvature tensor is defined as 
\begin{equation}
	R_{\rho\sigma}{}^\mu{}_{\nu} \equiv 2\, \partial_{[\rho} \Gamma^\mu_{\sigma]\nu} + 2 \, \Gamma_{\tau [\rho}^\mu \,\Gamma_{\sigma]\nu}^\tau, \label{Riemann} 
\end{equation}
where here and throughout the text we use symmetrization $()$ and antisymmetrization $[]$ with weight one, i.e. $(\mu\nu) = \nicefrac12(\mu\nu + \nu\mu)$, $[\mu\nu] = \nicefrac12(\mu\nu -\nu\mu)$ etc.

The Ricci tensor, the Ricci scalar and the Einstein-Hilbert action are given by
\begin{eqnarray}	
R_{\mu\nu} &\equiv & 	R_{\rho\mu}{}^\rho{}_{\nu} ,\\
R&\equiv & R_{\rho\mu}{}^\rho{}_{\nu} g^{\mu\nu} = R_{\mu\nu} g^{\mu\nu},\\	
S_{EH} &=& \frac{M_{P}^{2}}{2}\int d^4x\,\sqrt{-g}\, R=\frac{1}{16 \pi G_N} \int d^4x\,\sqrt{-g}\, R, \label{EHaction} 
\end{eqnarray}
where $g\equiv \textrm{det}(g_{\mu\nu})$, $G_N$ denotes Newton's constant, and $M_{P}$ is the reduced Planck mass (with $c=1$, $\hbar=1$),
\begin{equation}
	M_P = \frac{1}{\sqrt{8 \pi G_N}} 
	= 2.44 \cdot 10^{18}\, \textrm{GeV}.
\label{PlanckMass} 
\end{equation}

\subsection{Vierbein and Cartan's formalism}
\label{subsec:Vierbein}
In order to describe the coupling to fermions, an orthonormal basis of tangent vectors,  $e_a$ ($a=0,1,2,3$), is introduced at each point of the spacetime manifold,
\begin{equation}
g(e_a,e_b)=\eta_{ab},	
\end{equation}
where 
\begin{equation}
\eta_{ab}=\textrm{diag}(-1,+1,+1,+1)	
\end{equation} 
is the flat Minkowski metric. Orthonormality fixes the $e_a$ only up to local Lorentz transformations, $e_a\rightarrow e_b\, \Lambda^b{}_{a}(x)$, with $\Lambda^b{}_a(x) \in SO(1,3)$. 

The basis change from $e_a$ to the coordinate basis vectors $\partial_{\mu}$ and vice versa
 is written as
\begin{equation}
e_a=e_a^\mu (x) \partial_\mu, \qquad \partial_\mu = e_{\mu}^a (x) e_a, 
\end{equation}
where  $e_\mu^a e_a^\nu=\delta_\mu^\nu$, $e_a^\mu e_\mu^b =\delta_a^b$. By a common abuse of terminology, we refer to both the vectors $e_a$ as well as the matrix of conversion coefficients, $e_\mu^a(x)$, as the \emph{vierbein}.

The vierbein $e_{\mu}^{a}$ and its inverse can be used to convert the ``curved'' (or ``world'') indices $\mu,\nu,\ldots$ of any tensor field to ``flat'' (or ``local Lorentz'') indices $a,b,\ldots$, 
in particular,  
\begin{equation}
	g_{\mu\nu}(x) = e_\mu^a(x) e_\nu^b(x) \eta_{ab},
\label{equivalence} 
\end{equation}
which  also implies  
\begin{equation}
	\sqrt{-g} \equiv \sqrt{-\det (g_{\mu\nu})}=  {\rm det}\, e_\mu^a \equiv e
\end{equation}
and shows that the vierbein $e_{\mu}^{a}$ contains the same information as the metric.

When tensor fields are expressed in terms of the flat indices $a,b,\ldots$, the connection coefficients of the covariant derivative are denoted as $\omega_{\mu}{}^a{}_b$, so that, e.g., $\nabla_{\mu} V^a=\partial_{\mu}V^a+\omega_{\mu}{}^{a}{}_{b}V^b$. This covariant derivative is equivalent to the one for curved indices as defined by the Christoffel symbols (in the sense that $\nabla_{\mu}V^a=e_{\nu}^a\nabla_{\mu}V^\nu$ etc.) if
\begin{equation}
	\nabla_\mu e_\nu^a \equiv 
	\partial_\mu e_\nu^a + \omega_\mu{}^a{}_b e^b_\nu - \Gamma_{\mu\nu}^\rho e_\rho^a =0,
\label{la1} 
\end{equation}
where, as suggested by the notation, the left hand side can be viewed as a total covariant derivative that acts on curved and flat indices of the vierbein, implementing covariance  with respect to general coordinate and  local Lorentz transformations.  
Eq.~(\ref{la1}) is sometimes referred to as the ``vierbein postulate''.

In supergravity, it is useful to define also a derivative operator, $D_\mu$, that is covariant only with respect to local Lorentz transformations, but not necessarily 
with respect to general coordinate transformations, i.e., all local Lorentz indices
will be contracted with spin connections, but there are no Christoffel symbols
that contract any world index. 
An important example is the Lorentz covariant derivative of the vierbein $e^a_\nu$,
\begin{equation}
D_\mu e^a_\nu \equiv \partial_\mu e^a_\nu + \omega_{\mu}{}^{a}{}_b \, e^b_{\nu}.
\end{equation}
In view of (\ref{la1}), this expression does not vanish, but is equal to $\Gamma_{\mu\nu}^\rho e_\rho^a$.
Recalling that the antisymmetrization $\Gamma_{[\mu\nu]}^{\rho}=\frac{1}{2} T_{\mu\nu}{}^{\rho}$ is just the torsion tensor of this connection, we can thus write 
\begin{equation}
D_{[\mu} e^a_{\nu]} =\frac{1}{2} T_{\mu\nu}{}^{a} = \frac{1}{2} T_{\mu\nu}{}^{\rho}e_{\rho}^a,\label{antisymtorsion}
\end{equation}
which is now a proper tensor field that vanishes for the Levi-Civita connection.

Just as for the Levi--Civita connection in curved indices, the vanishing of the torsion tensor can be used to deduce an explicit expression for the spin connection
\index{spin connection} in terms of the vierbein, $\omega_\mu{}^{ab} = \omega_\mu{}^{ab}(e)$.
To this end, one considers 
\begin{equation}
	t^{dc,a} \equiv e^{\nu[d} e^{\mu c]}\left(
	\partial_\mu e_\nu^a + \omega_\mu{}^a{}_b e^b_\nu\right), \nonumber	
\end{equation} 
which is is zero because of the assumed vanishing torsion. This then also implies the vanishing of the following combination,
\begin{equation}
	t^{dc,a} - t^{ca,d} - t^{ad,c} = 0,
\label{sumt3} 
\end{equation}
in which only one term in the spin connection, $e^{\rho a}\omega_\rho{}^{cd}$, survives so that, after multiplying by $e_{\mu a}$, one obtains
\begin{equation}
	\omega_\mu{}^{cd} (e) = 2 e^{\nu[c} 
	\partial_{[\mu}e_{\nu]}^{d]} - e_{a \mu} e^{\nu[c}e^{\sigma d]}
	\partial_\nu e_\sigma^a. \label{spinconnection} 
\end{equation}
This connection is called the \emph{torsion-free spin connection}, whose equivalence to
the Levi Civita connection   (\ref{LeviCivita}) can also be verified directly. 

As mentioned above, even though the Lorentz covariant derivative, $D_\mu$, is not covariant with respect to general coordinate transformations, the \emph{antisymmetrized} derivative (\ref{antisymtorsion}) still forms a proper tensor field. 
In supergravity, all equations can similarly be expressed in terms of antisymmetrized Lorentz covariant derivatives only so that the Christoffel symbols $\Gamma_{\mu\nu}^{\rho}$ never occur. 
This allows one to use compact differential form notation, which can substantially reduce the index clutter in computations. 
To this end, we introduce the co-frame of one-forms, $e^a$, dual to the vectors $e_b$,
\begin{equation}
e^a \equiv e^a_\mu dx^\mu, 
\end{equation}
as well as the connection one-forms, 
\begin{equation}
\omega^{a}{}_{b}\equiv \omega_{\mu}{}^{a}{}_{b} dx^\mu.	
\end{equation}

In terms of $e^a$, the expression (\ref{antisymtorsion}) for the torsion tensor\index{torsion tensor} reads  
\begin{equation}
	D e^a \equiv de^a + \omega^a{}_b \wedge e^b = T^a = \frac12 dx^\mu \wedge dx^\nu T_{\mu\nu}{}^a, \label{spin-connection} 
\end{equation}
whereas the curvature tensor of the connection $\omega_\mu{}^b{}_a$ is given by the two-form
\begin{equation}
	R^a{}_b = d \omega^a{}_b + \omega^a{}_c \wedge \omega^c{}_b.
\label{Riemannform} 
\end{equation}

The conditions for the  connection $\omega_\mu{}^b{}_a$ to be equivalent to the Levi--Civita connection can now be written as
\begin{itemize}
	\setlength\itemsep{2mm}
	\item $D \eta_{ab}\equiv d \eta_{ab} + \omega_a{}^c \eta_{cb} + \omega_b{}^c \eta_{ac} = 0$ (metric compatibility); 
	\item $T^a = 0$ (vanishing torsion).
\end{itemize}
The first condition implies that the spin connection is antisymmetric in its indices, $\omega^{(ab)} = 0$, when both are raised or lowered with $\eta$, and  
the second equation (in combination with (\ref{spin-connection})) is equivalent to
\begin{equation}
		D  e^a =0.
\end{equation}

In Cartan's formalism the Einstein--Hilbert action \index{Einstein--Hilbert action}reads
\begin{equation}
	S_{EH} = \frac{M_P^2}{4}\int R^{ab}\wedge e^c \wedge e^d \epsilon_{abcd}, \label{EinsteinHilbertform} 
\end{equation}
where $\epsilon_{0123} = 1$, and 
the standard form (\ref{EHaction}) is recovered via  the identification of the 
four-dimensional measure,
\begin{equation}
		dx^\mu \wedge dx^\nu \wedge dx^\rho \wedge dx^\sigma = - d^4 x \, e \, \epsilon^{\mu \nu \rho \sigma},
\end{equation}
with $\epsilon_{\mu \nu \rho \sigma} = e_\mu^a e_\nu^b e_\rho^c e_\sigma^d \, \epsilon_{abcd}$. 

Considering the vierbein and spin connection as independent fields, yields the field equations

\begin{eqnarray}
	\frac{\delta S}{\delta e_\mu^a} = 0 & \Leftrightarrow & G_{\mu\nu} = 0, \label{eom1} \\[2mm]
	\frac{\delta S}{\delta \omega_\mu^{ab}} =0& \Leftrightarrow & T^a = 0,
\label{eom2} 
\end{eqnarray}
where
\begin{equation}
	G_{\mu\nu} \equiv R_{\mu\nu} - \frac12 g_{\mu\nu} R \label{Einsteincombo} 
\end{equation}
is the usual Einstein-tensor. Hence, in this formalism $\omega_{\mu}{}^{ab}$ is dynamically fixed to be the torsion-free spin connection, and one recovers the standard Einstein equations.

In supergravity, on the other hand, the spin connection also appears in the kinetic term of the gravitino so that Eq. (\ref{eom2}) receives a correction term bilinear in the gravitino fields. Thus, in supergravity, the spin connection as determined from the field equation has torsion and is no longer equivalent to the torsion-free Levi-Civita connection. This is not a problem,  because the Christoffel symbols never appear in the theory, and one is always free to redefine the spin connection as
\begin{equation}
	\omega_{\mu}{}^{ab}=\omega_{\mu}{}^{ab}(e) + \kappa_{\mu}{}^{ab},
\end{equation}
where $\omega_{\mu}{}^{ab}(e)$ is the torsion-free spin connection (\ref{spinconnection}), and $\kappa_{\mu}{}^{ab}$ is the \emph{contorsion tensor},
\begin{equation}
\kappa_\mu{}^{ab} = e_\mu^c\left(T^{ab}{}_c - T_c{}^{a\,b}- T^b{}_c{}^a\right), \label{contorsion} 
\end{equation}
bilinear in the fermions. Espressing the theory in terms of $\omega_{\mu}{}^{ab}(e)$ then introduces explicit four-Fermion interactions in the action.


\subsection{Gamma matrices}
We first introduce gamma matrices in flat Minkowski space with orthonormal basis vectors labelled by the indices $(a,b,\ldots = 0,1,2,3)$ and the standard  Minkowski metric
\begin{equation}
\eta_{ab}=\textrm{diag}(-1,+1,+1,+1)	.
\end{equation} 
The Lorentz algebra generators are denoted by $M_{ab}=-M_{ba}$ and satisfy
 \begin{equation}
 	[M_{ab},M_{cd}] = - 2\,\eta_{c[a} M_{b]d} +2\, \eta_{d[a}M_{b]c} . \label{Malgebrasmall}
 \end{equation}

The gamma matrices, $\gamma_a$, obey following basic relations 
\begin{eqnarray}
	\label{Clifford}
	&&\left\{\gamma_a, \gamma_b\right\} = 2 \,\eta_{ab} \,{\mathbbm 1}_{4}, \\
	&&\gamma_{0}^{\dagger}=-\gamma_{0},\quad  
	\gamma_{i}^{\dagger}=+\gamma_{i},\quad
	\gamma_{a}^{T}=\pm\gamma_{a}, \\
	&&\label{Sigmadefinition} \Sigma_{ab}:=\frac{1}{4}[\gamma_{a},\gamma_{b}], \\
	&&\gamma_{a_{1}\ldots a_{p}}\equiv\gamma_{[a_{1}}\gamma_{a_{2}}\ldots \gamma_{a_{p}]} , \\
	&&	\label{gamm5} 	\gamma_{5}\equiv \gamma^5 \equiv -i \gamma^0 \gamma^1 \gamma^2 \gamma^3 = +i \gamma_0 \gamma_1 \gamma_2 \gamma_3, \\
	&& 	(\gamma_{5})^{2} = {\mathbbm 1}_{4}, \quad 
	\{\gamma_{5},\gamma_{a}\}=0, \quad  [\gamma_{5},\Sigma_{ab}]=0.  
\end{eqnarray}
We also introduce the charge conjugation matrix $C$, satisfying
\begin{eqnarray}
   	&& 	C^T = - C = C^{-1} =C^{\dagger}, \\
   	&& \gamma_a^T = - C \gamma_a C^{-1},
\end{eqnarray}
implying the following symmetry properties 
\begin{equation}
	\begin{array}{lclcl}
		C^T = -C, \qquad  (C \gamma^{a})^T = (C \gamma^{a}), \qquad (C \gamma^{ab})^T = (C \gamma^{ab}), \\[3mm]
		(C \gamma^{abc})^T = - (C \gamma^{abc}),  \qquad  (C \gamma^{abcd})^T = -(C \gamma^{abcd}).
	\end{array}
\end{equation}
The matrix $\gamma_5$ also enters a number of useful duality relations between the antisymmetrized products of gamma matrices,
\begin{equation}
   		\begin{array}{l}
		\displaystyle \gamma^{abc} = i\, \epsilon^{abcd} \gamma_d \gamma_5, \qquad i \,\gamma_a \gamma_5 = \frac{1}{3!} \epsilon_{abcd}\gamma^{bcd},\\[3mm]
		\displaystyle \gamma^{abcd} = - i\, \epsilon^{abcd} \gamma_5, \qquad i \,\gamma_5 =  \frac{1}{4!}\epsilon_{abcd}\gamma^{abcd}, \\[3mm]
		\displaystyle \gamma^{ab} = \frac{i}{2} \epsilon^{abcd} \gamma_{cd} \gamma_5,
	\end{array}
   \end{equation}
where $\epsilon_{abcd}$ is the antisymmetric epsilon tensor with 
\begin{equation}
\epsilon_{0123} = 1.
\end{equation}

\subsection{Spinors}
Throughout this text, we use the four-component spinor notation. A translation to the two-component spinor notation in our conventions can be found in \cite{DallAgata:2021uvl}.
\subsubsection{Dirac spinors}
Dirac spinors are elements of the representation space of the Clifford algebra (\ref{Clifford}). As the
 matrices $\Sigma_{ab}$ provide an explicit linear representation of the Lorentz algebra generators $M_{ab}$, Dirac spinors inherit also a Lorentz algebra representation.

For a generic Dirac spinor, $\psi$, we define the \emph{Dirac conjugate} as
\index{Dirac conjugate}
\begin{equation}
	\overline{\psi} \equiv i \psi^\dagger \gamma^0 = -i\psi^\dagger \gamma_0
\label{Diracconj} 
\end{equation}
so that bilinears such as $\overline{\psi}\chi$ are Lorentz invariant, because of $\Sigma_{ab}^{\dagger}\gamma_{0}=-\gamma_{0}\Sigma_{ab} $.

On 4D Dirac spinors, however, the Lorentz algebra representation provided by $\Sigma_{ab}$ is reducible.
This reducibility can be resolved either by a chirality condition, leading to Weyl spinors, or by a reality condition, leading to Majorana spinors.

\subsubsection{The Weyl condition} 

The Weyl condition projects out the part of a spinor that has a particular handedness.
In order to impose it, one uses the chirality projectors 
\begin{equation}
	P_L \equiv \frac12 \left(1 + \gamma^5\right), \quad P_R \equiv \frac12 \left(1 - \gamma^5\right), \label{chirality} 
\end{equation}
and defines left- and right-handed spinors,
\begin{equation}
	\psi_L \equiv P_L \psi, \qquad \psi_R \equiv P_R \psi, \qquad \textrm{(Weyl condition)}
\label{LRrep} 
\end{equation}
satisfying  $\gamma^5 \psi_L = \psi_L$ and $\gamma^5 \psi_R = -\psi_R$. 
This projection is consistent with Lorentz covariance, and left- and right-handed spinors form separate representations of the Lorentz group.

Using $\gamma_{5}^{\dagger}=\gamma_{5}$ as well as  $P_L \gamma_0 = \gamma_0 P_R$ and $P_R \gamma_0 = \gamma_0 P_L$, one finds
\begin{equation}
	\overline{\psi_R} = \overline \psi P_L, \qquad  \overline{\psi_L} = \overline \psi P_R, \label{conjugations} 
\end{equation}
where 
\begin{equation}
	\overline{\psi_R} \equiv \overline{(P_R \psi)} = -i (P_R \psi)^\dagger \gamma_0.
\end{equation}
Because of this we will often write 
\begin{equation}
	\overline \psi_L \equiv \overline{\psi} P_L = \overline{\psi_R}, \qquad \overline \psi_R \equiv \overline{\psi} P_R = \overline{\psi_L} .
\label{defalte} 
\end{equation}

\subsubsection{The Majorana condition} 

The Majorana condition is a reality condition that is usually expressed in terms of the charge conjugation matrix, $C$.
In terms of $C$, the charge conjugate spinor of a four-component spinor, $\psi$, is defined as  
\begin{equation}
	\psi^c = C \overline{\psi}^T = i C \gamma^{0 T} \psi^* . \label{ccspinor} 
\end{equation}
A \emph{Majorana spinor} is then a spinor that equals its own charge conjugate, 
\index{Majorana spinor} 
\begin{equation}
	\psi^c = \psi. \qquad \textrm{(Majorana condition)}
\label{Majodef} 
\end{equation}

For a Majorana spinor, the Dirac conjugate (\ref{Diracconj}) takes on the simple form 
\begin{equation}
	\overline{\psi} = \psi^T C.
\label{DconMa} 
\end{equation}

For \emph{anti-}commuting Majorana spinors, this then implies 
\begin{equation}\label{trasporules}
	{\overline{\psi}}_{1}M\psi_{2} = \left\{ 
	\begin{array}{l}
		+{\overline{\psi}}_{2}M\psi_{1} \qquad \textrm{for } M={\mathbbm 1}_{4}, \gamma_{abc}, \gamma_{abcd}\\[4mm]
		-{\overline{\psi}}_{2}M\psi_{1} \qquad \textrm{for } M=\gamma_{a}, \gamma_{ab} 
	\end{array}
	\right.
\end{equation}

\emph{Unless stated otherwise, we will always use anti-commuting Majorana spinors, but often also take in addition the chiral projections $\psi_L$ and $\psi_R$ of these Majorana spinors, which therefore are \textbf{not} independent.} 

More specifically, the chiral projections, $\psi_L$ and $\psi_R$, of a Majorana spinor $\psi$ satisfy 
\begin{equation}
	(\psi_L)^c = \psi_R, \qquad (\psi_R)^c = \psi_L,
\label{dependencepsi2} 
\end{equation}
showing that $\psi_L$ and $\psi_{R}$ are no longer  Majorana spinors themselves. In other words, a spinor in 4D cannot be simultaneously chiral and Majorana, but it nevertheless does make sense to talk about the chiral projections $\psi_{L}$ or $\psi_{R}$ of a given Majorana spinor $\psi$.

In fact, several identities that hold for Majorana spinors still have a close analogue 
for their chiral projections. In particular, the chiral projection $\psi_L$ of a Majorana spinor $\psi$ satisfies
\begin{equation}
\overline{\psi}_{L}\equiv\overline{\psi}P_{L}	
=(\psi_{L})^{T}C \label{MajoranaconjugateLR}
\end{equation}
and similarly for $\psi_R$, as follows from (\ref{DconMa}) and $C\gamma_5=\gamma_5^TC$.
From this,  one can obtain more symmetry properties for the chiral projections that are very similar to eqs. (\ref{trasporules}) for the Majorana spinors themselves,

\begin{equation}
	\begin{array}{l}
		\overline{\chi}_L \psi_L = \chi_L^T C\psi_L =- \psi_L^T C^T \chi_L = \overline{\psi}_L \chi_L, \\[2mm]
		\overline \chi_L \gamma^a \psi_R = - \overline \psi_R \gamma^a \chi_L, \quad \overline \chi_L \gamma^{ab} \psi_L = - \overline \psi_L \gamma^{ab} \chi_L \\[2mm]
		\overline \chi_L \gamma^{abc} \psi_R = \overline \psi_R \gamma^{abc} \chi_L.
	\end{array}\label{bispinors}
\end{equation}

\subsection{Spinors in curved spacetime}
\label{subsec:Spinorscurved}
Spinors are double-valued representations of the Lorentz group. On a curved manifold, the relevant Lorentz transformations are the local Lorentz transformations of the vierbein so that
\begin{equation}
D_{\mu}\psi = \partial_{\mu} \psi + \frac{1}{4}\omega_{\mu}{}^{ab}\gamma_{ab} \psi,
\label{covariantDspinor}
\end{equation}
is the proper Lorentz covariant derivative of a fermion  field $\psi$. 

The gamma matrices with a curved index $\mu$ are obtained from the
constant $\gamma_{a}$ via contaction with a vierbein: 
\begin{equation}
	\gamma_{\mu}\equiv e_{\mu}^{a}\gamma_{a}.
\end{equation}
The $\gamma_{\mu}$ are then in general no longer constant, and they transform non-trivially under a variation of the vierbein.

\subsection{Fierz identities} 

We will often need to rewrite 3 or 4-Fermi terms to complete our analysis of the supersymmetry properties of an action, and hence Fierz identities
\index{Fierz identities} will be extremely useful.
We list here the main ones for two spinors: 
\begin{eqnarray}
	\psi_R \overline{\chi}_R &=& -\frac12 \overline\chi_R \psi_R\, \, P_R + \frac18 \overline{\chi}_R \gamma_{ab}\psi_R \; \gamma^{ab}\, P_R, \label{FierzRR} \\[2mm]
	\psi_R \overline{\chi}_L &=& -\frac12 \overline{\chi}_L \gamma^a \psi_R \; \gamma_a\, P_L, \label{FierzRL} 
\end{eqnarray}
where for the sake of clarity we explicitly left the projectors on the right hand side.

We often make use of spinor 1-forms $\psi = dx^\mu \psi_\mu$. 
Exchanging such spinor one-forms then leads to an additional minus sign from the anti-commutativity of the wedge product, and hence one has
\begin{eqnarray}
	\psi_R \wedge \overline{\psi}_R &=& - \frac18 \overline{\psi}_R \wedge\gamma_{ab}\psi_R \; \gamma^{ab}\, P_R, \\[2mm]
	\psi_R \wedge \overline{\psi}_L &=& \frac12 \overline{\psi}_L \wedge  \gamma^a \psi_R \; \gamma_a\, P_L, 
\end{eqnarray}
where now $\overline{\psi}_L \wedge \psi_L = 0$ because of (\ref{bispinors}) and the wedge product.
A crucial consequence is the so-called cyclic identity:
\begin{equation}
	\gamma^a \psi_L \wedge \overline \psi_L \wedge \gamma_a \psi_R = 0.
\label{cyclicid} 
\end{equation}

\subsection{Lie derivative on p-forms.}
\label{subsec:Liederivative}

The Lie derivative of a p-form $A_p$ along the flow of a vector field $V$ is defined as
	$$ L_V A_p = \lim_{t\to 0} \frac1t \left(\sigma_{t}^* A_p(\sigma_{t}(x)) - A_p(x)\right),$$
	where $\sigma_{t}^*$ is the pull back of the differential form along the flow generated by the vector field $V$.
	When applied to a scalar valued $p$-form this reduces to
	$$ L_V A_p = (\imath_V d + d \imath_V)A_p.$$

\subsection{The supersymmetry algebra} 
\label{sub:the_supersymmetry_algebra}

The $\mathcal{N}=1$ supersymmetry algebra in 4 dimensions can be represented as follows
\begin{equation}
	\begin{array}{rcl}
	\{Q, \overline{Q}\} &=& - 2 i \gamma^a {\cal P}_a, \\[2mm]
	[{\cal P}_a, Q] &=& 0, \\[2mm]
	[{\cal M}_{ab},Q] &=& \frac{i}{2}\,\gamma_{ab} Q, \\[2mm]
	[R, Q] &=& i \gamma_5 Q, \\[2mm]
	[{\cal P}_a, {\cal P}_b] &=& 0,\\[2mm]
	[{\cal P}_a, {\cal M}_{bc}] &=& - 2i \eta_{a[b} {\cal P}_{c]}, \\[2mm]
	[{\cal M}_{ab}, {\cal M}_{cd}] &=& 2 i \eta_{c[a} {\cal M}_{b]d}- 2i\eta_{d[a} {\cal M}_{b]c},
	\end{array}
	\label{susyalgebra}
\end{equation}
where we used Hermitian generators ${\cal P}_a$ and ${\cal M}_{ab}$ for the Poincar\'{e} algebra.
Just as in (\ref{Malgebrasmall}), we sometimes also use their anti-Hermitian counterparts, 
\begin{equation}
		P_a = i\, {\cal P}_a, \qquad M_{ab} = i\, {\cal M}_{ab}
\end{equation}
when this is more convenient. 

\end{document}